\documentclass[a4paper,11pt]{article}
\pdfoutput=1 % if your are submitting a pdflatex (i.e. if you have
\usepackage{jheppub} % for details on the use of the package, please
\usepackage[T1]{fontenc} % if needed
\usepackage{xspace}
\usepackage{mciteplus}
\usepackage{hyperref}    % Hyperlinks in references                                                                     
\usepackage{upgreek}

\usepackage[all]{hypcap} % Internal hyperlinks to floats.    

\def\micron  {\ensuremath{{\,\upmu\rm m}}\xspace}

\def\figwidth{0.45}
\def\widefigwidth{0.45}
\def\CP                {{\ensuremath{C\!P}}\xspace}
\def\Apm{\ensuremath{A_{\pm}}\xspace}
\def\BsToKMuNu{\ensuremath{B_s^0 \to K^- \mu^+ \nu_{\mu}}\xspace}
\def\BsToKstMuNu{\ensuremath{B_s^0 \to K^{\ast-} \mu^+ \nu_{\mu}}\xspace}
\def\BsToDsMuNu{\ensuremath{B_s^0 \to D_s^- \mu^+ \nu_{\mu}}\xspace}
\def\BsbToDsMuNu{\ensuremath{\bar{B}_s^0 \to D_s^+ \mu^- \bar{\nu}_{\mu}}\xspace}
\def\BsToDstMuNu{\ensuremath{B_s^0 \to D_s^{\ast-} \mu^+ \nu_{\mu}}\xspace}
\def\LbToPMuNu{\ensuremath{\Lambda_b^0 \to p \mu^- \bar{\nu}_{\mu}}\xspace}
\def\vecPMiss{\ensuremath{\vec{P}_{\rm miss}}\xspace}
\def\vecPVis{\ensuremath{\vec{P}_{\rm vis}}\xspace}
\def\PLMiss{\ensuremath{P_{\rm miss}^{\parallel}}\xspace}
\def\PLVis{\ensuremath{P_{\rm vis}^{\parallel}}\xspace}
\def\PTVis{\ensuremath{P_{\rm vis}^{\perp}}\xspace}
\def\PTMiss{\ensuremath{P_{\rm miss}^{\perp}}\xspace}

\def\Pinf{\ensuremath{P_{\rm inf}}\xspace}
\def\MVis{\ensuremath{M_{\rm vis}}\xspace}
\def\MCorr{\ensuremath{M_{\rm corr}}\xspace}
\def\EVis{\ensuremath{E_{\rm vis}}\xspace}
\def\PinfTheta{\ensuremath{P_{\rm inf}^{\theta}}\xspace}
\def\Pinf{\ensuremath{P_{\rm inf}}\xspace}
\def\BsBsb{\ensuremath{B_s^0-\bar{B}_s^0}\xspace}
\def\BdBdb{\ensuremath{B_d^0-\bar{B}_d^0}\xspace}
\def\Bs{\ensuremath{B_s^0}\xspace}
\def\Bsb{\ensuremath{\bar{B}_s^0}\xspace}

\usepackage{ifthen}
\newboolean{articletitles}
\setboolean{articletitles}{true} % False removes titles in references                                         
\newboolean{uprightparticles}
\setboolean{uprightparticles}{false} %True for upright particle symbols                            
\newboolean{inbibliography}
\setboolean{inbibliography}{false} %True once you enter the bibliography                                  

\title{\boldmath Reconstruction of semileptonically decaying beauty hadrons produced in high energy $pp$ collisions}
\author[a]{G.~Ciezarek}
\author[b]{A.~Lupato}
\author[c]{M.~Rotondo}
\author[d,1]{M.~Vesterinen\note{Corresponding author.}}

\affiliation[a]{Nikhef, Netherlands} %One University,\\some-street, Country}
\affiliation[b]{INFN Padova, Italy} %Another University,\\different-address, Country}
\affiliation[c]{Laboratori Nazionali di Frascati, Italy}
\affiliation[d]{Physicalisches Institute Heidelberg, Germany} %A School for Advanced Studies,\\some-location, Country}

\emailAdd{greg.ciezarek@cern.ch}
\emailAdd{anna.lupato@cern.ch}
\emailAdd{marcello.rotondo@cern.ch}
\emailAdd{mika.vesterinen@cern.ch}

\abstract{It is well known that in $b$-hadron decays with a single unreconstructible final state particle,
the decay kinematics can be solved up to a quadratic ambiguity,
without any knowledge of the $b$-hadron momentum.
We present a method to infer the momenta of $b$-hadrons produced in hadron collider experiments
using information from their reconstructed flight vectors.
Our method is strictly agnostic to the decay itself, which implies
that it can be validated with control samples of topologically similar decays to fully reconstructible final states.
A multivariate regression algorithm based on the flight information
provides a $b$-hadron momentum estimate with a resolution of around 60\%
which is sufficient to select the correct solution to the quadratic equation in around 70\% of cases.
This will improve the ability of hadron collider experiments to make differential
decay rate measurements with semileptonic $b$-hadron decays.}

\begin{document} 
\maketitle
\flushbottom

%%%%%%%%%%%%%%%%%%%%%%%%%%%%%%%%%%%%%%%%%%%%%%%%%%%%%%%%%%%%%%%%%%%%%%
\section{Introduction} \label{sec:Introduction}
%%%%%%%%%%%%%%%%%%%%%%%%%%%%%%%%%%%%%%%%%%%%%%%%%%%%%%%%%%%%%%%%%%%%%%

The study of semileptonic decays of beauty hadrons, with transitions $b \to \ell\nu q_u$ ($q_u = c,u$), 
is of great interest since these decays offer a theoretically clean determination of the magnitudes of 
the CKM matrix elements $V_{ub}$ and $V_{cb}$.
The majority of studies have been restricted to $B^+$ and $B^0$ mesons produced by $e^+e^-$ colliders operating
at the $\Upsilon(4S)$ resonance.
The presence of an unreconstructible neutrino in the final state poses an experimental challenge.
However, in exclusive production of $B\bar{B}$ meson pairs in $e^+e^-$ collisions at the $\Upsilon(4S)$ resonance,
the decay kinematics of the $B$ can be resolved by balancing against the $\bar{B}$ decay or vice versa.

Hadron collider experiments offer an enticing opportunity to make complementary studies
with other $b$-hadron species.
The busy hadronic environment and inclusive production mechanism make these studies challenging.
However, there is an important advantage, which is that the $b$-hadrons tend to have a
large Lorentz boost, especially at the forward rapidities that are covered by the LHCb experiment~\cite{Alves:2008zz}.
The measured flight vector joining the primary $pp$ interaction vertex and the $b$-hadron decay vertex
can be exploited to constrain the decay kinematics~\cite{Dambach:2006ha}.
Decays with a single missing particle obviously have an unknown 3-momentum if an assumption
is made about the mass of this particle.
Two independent constraints are provided by momentum conservation transverse to the flight vector.
A third is provided by the assumption of the parent $b$-hadron mass,
though this constraint is quadratic and therefore presents two solutions.
One possibility~\cite{Stone:2014mza} is to consider $b$-hadrons that originate 
from decays of narrow excited $b$-hadron states.
If the other decay products from these decays are reconstructed then 
the mass of the excited state provides a further constraint
on the kinematics of the child $b$-hadron.

Recently, LHCb made the first observation of the decay \LbToPMuNu~\cite{Aaij:2015bfa},
and subsequently measured the ratio $|V_{ub}|/|V_{cb}|$, exploiting lattice QCD  calculations of the form factors~\cite{Detmold:2015aaa}.
A similar measurement with \BsToKMuNu is highly anticipated, given the precise lattice calculations~\cite{Flynn:2015mha}, which could permit the single most precise determination of $|V_{ub}|/|V_{cb}|$.
One of the challenges for a hadron collider experiment is to determine the 
invariant mass squared of the $\ell\nu$ system, denoted $q^2$, to permit a measurement of the differential decay rate as a function of this quantity.
As a result, the LHCb measurement~\cite{Aaij:2015bfa} of the \LbToPMuNu decay rate is restricted to a single high $q^2$ region\footnote{It should be noted that this is anyway the region in which the LQCD predictions~\cite{Flynn:2015mha} are most precise.}.
In order to suppress the contamination from decays originating outside this $q^2$ region, it was required that {\em both} solutions of the quadratic equation described above fall into the desired window.
The work presented here should permit similar studies with improved efficiency and granularity in $q^2$.

The idea is to identify variables that are correlated with the $b$-hadron momentum, but that are independent of the manner in which the $b$-hadron decays.
This implies that the method can be accurately validated with fully reconstructible decays that 
have a similar topology to the signal.
A regression based estimate of the $b$-hadron momentum, using these variables as input, can then be used to lift the quadratic ambiguity.
The studies presented here use the example of the LHCb experiment, but the ideas should be applicable to
any other current or future hadron collider experiment and several centre-of-mass energies are therefore considered.

%%%%%%%%%%%%%%%%%%%%%%%%%%%%%%%%%%%%%%%%%%%%%%%%%%%%%%%%%%%%%%%%%%%%%%
\section{Simulation of inclusive beauty production}
\label{sec:Simulation}
\begin{figure}[tb]\centering
\includegraphics[width=\figwidth\linewidth]{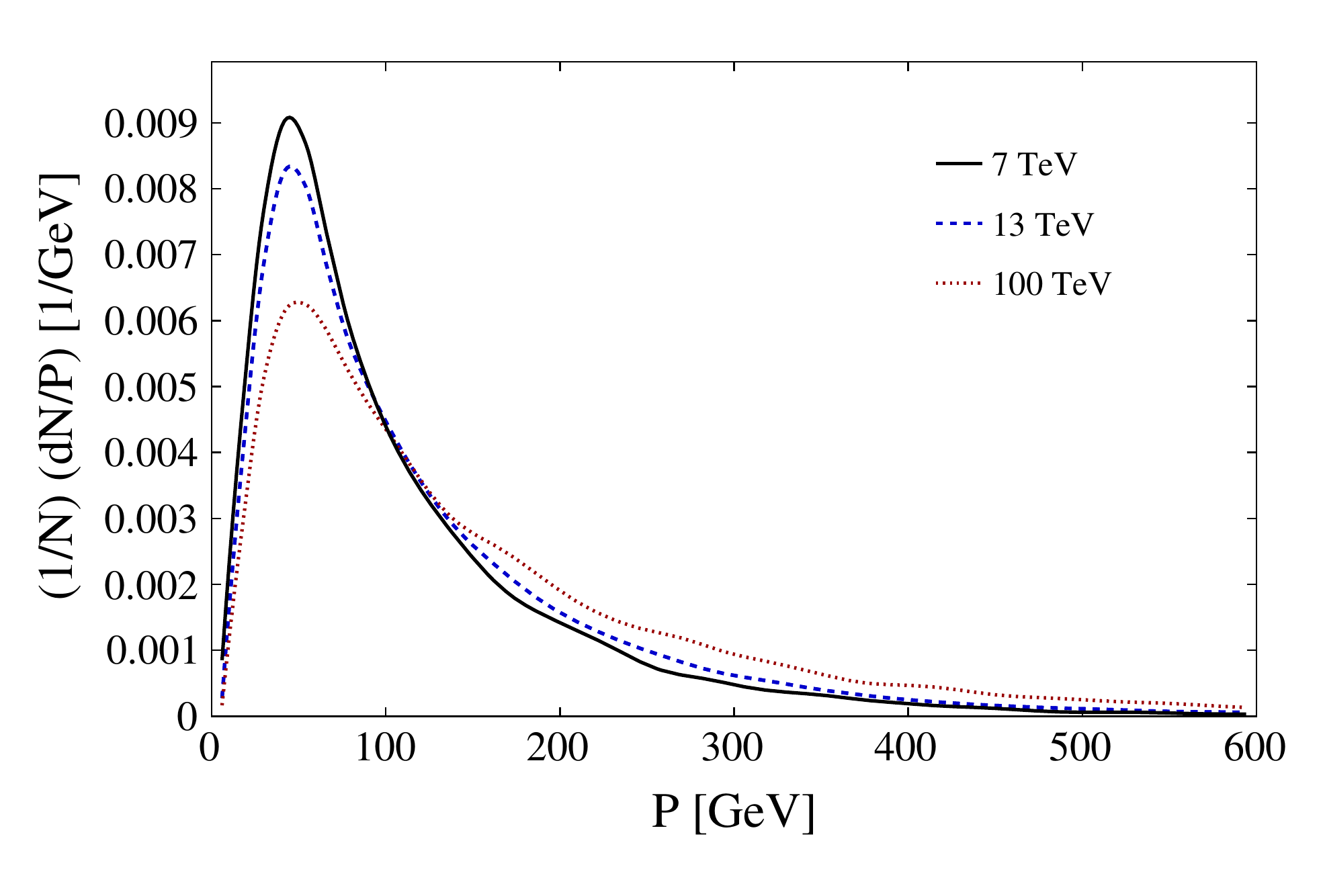}
\includegraphics[width=\figwidth\linewidth]{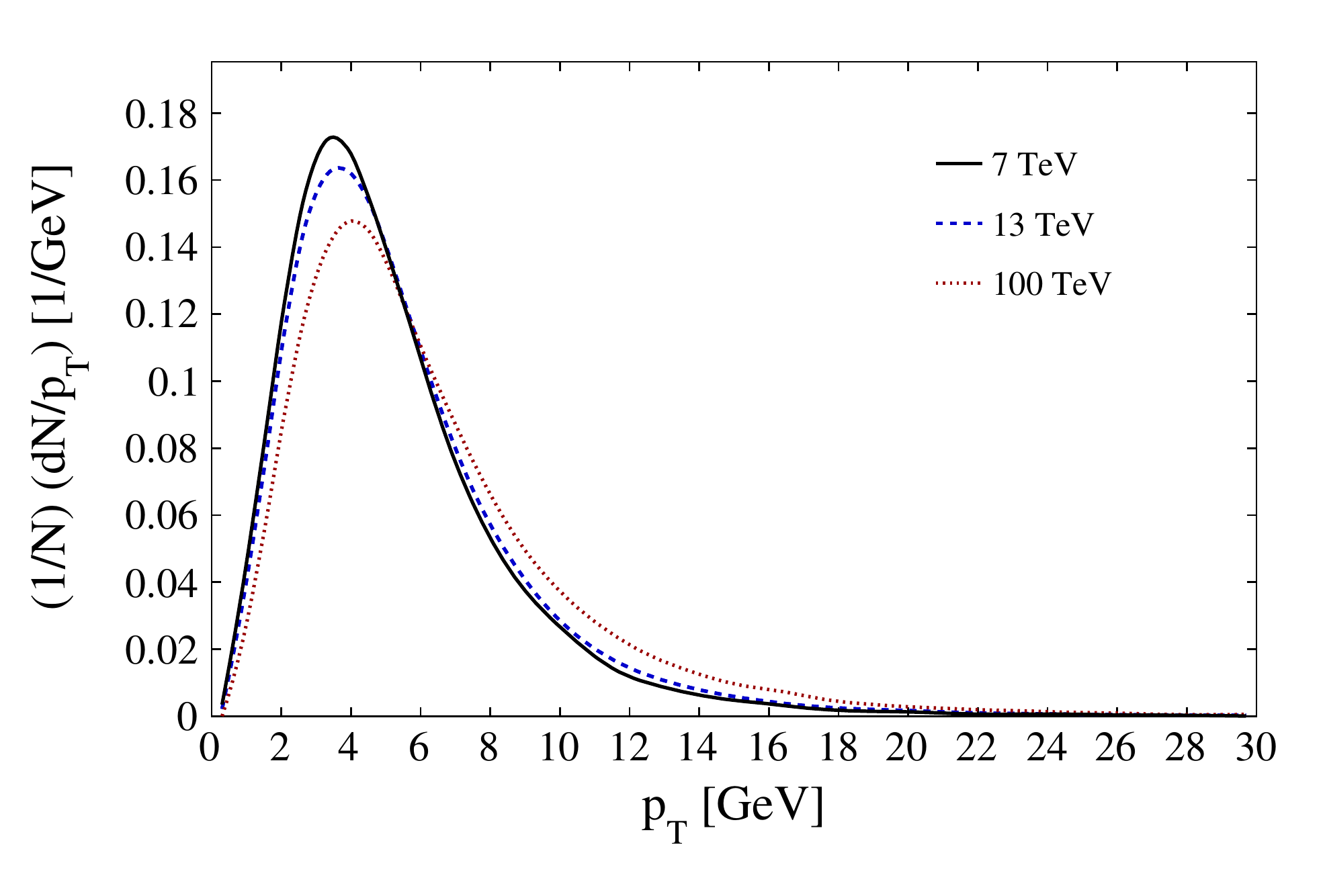}
\includegraphics[width=\figwidth\linewidth]{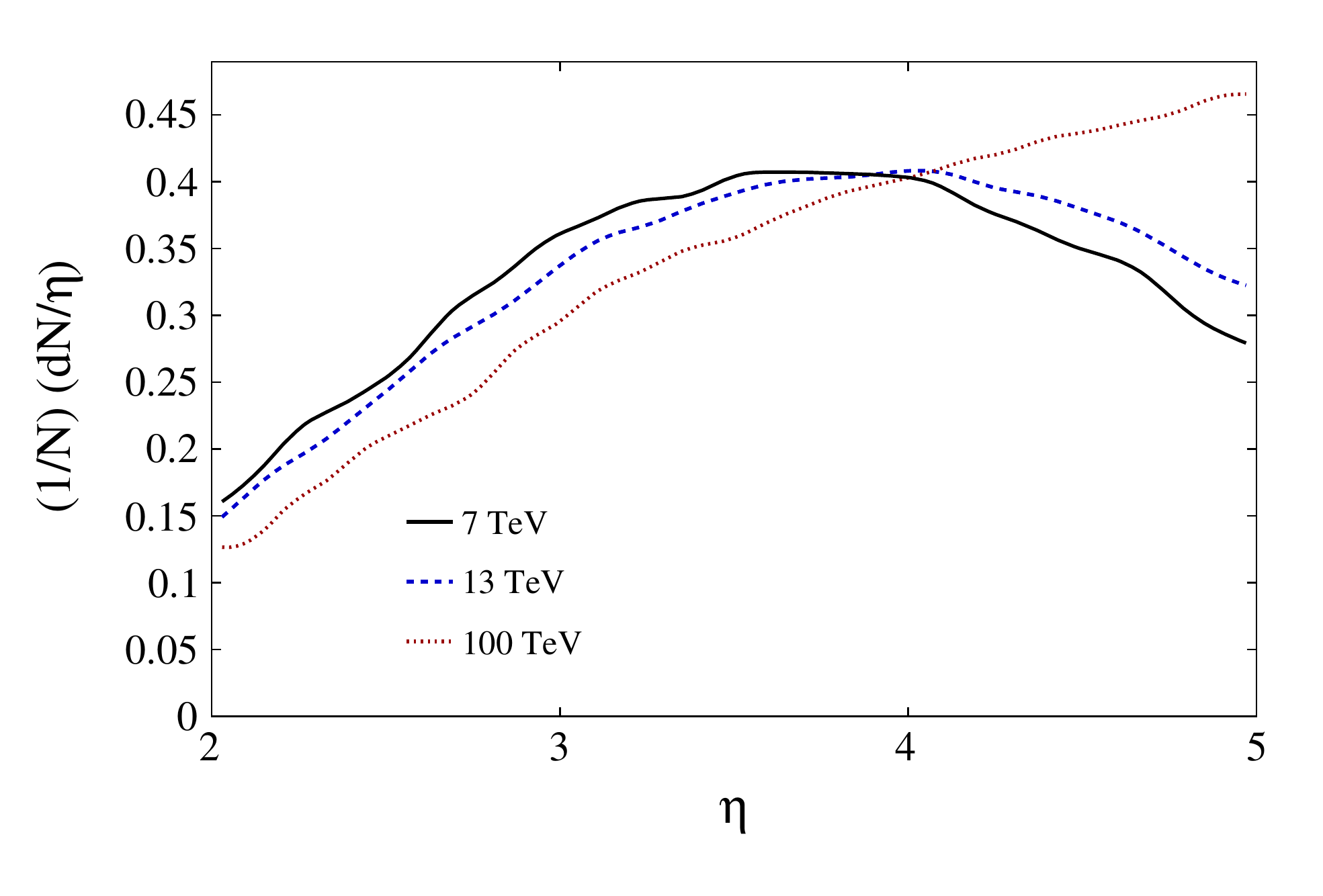}
\caption{\label{fig:B_NOCUTS}Basic $b$-hadron kinematic distributions
in our simulated event samples: (top left) momentum, (top right) transverse momentum, (lower) pseudorapidity.}
\end{figure}

%%%%%%%%%%%%%%%%%%%%%%%%%%%%%%%%%%%%%%%%%%%%%%%%%%%%%%%%%%%%%%%%%%%%%%
The Pythia~\cite{Sjostrand:2006za} event generator is used to simulate
inclusive $b$-hadron pair production in $pp$ collisions at three centre-of-mass energies, 
$\sqrt{s} = 7, 13, 100$~TeV.
Unless explicitly stated otherwise, the studies that follow
are based on the 13~TeV sample.
A right-handed coordinate system is defined with $z$ along the beam axis into the detector,
$y$ vertical and $x$ horizontal.
The magnitude of the momentum of a particle is denoted $P$, and the component transverse to the $z$
axis is defined as $p_T = P\sin\theta$.
The component of the momentum along the $z$ axis is denoted $p_z$.
A particle has a pseudorapidity defined as $\eta = -\ln\left(\tan(\theta/2)\right)$.
A particle of energy $E$ is defined to have a rapidity of $y = \frac{1}{2}\ln\left((E+p_z)/(E-p_z)\right)$.
Signal $b$-hadron candidates are required to be produced within the range $2 < \eta < 5$,
which corresponds to the approximate kinematic acceptance of the LHCb detector~\cite{LHCb-TDR-009}.
Fig.~\ref{fig:B_NOCUTS} shows, for each of the three centre-of-mass energies under consideration, 
the $P$, $p_T$ and $\eta$ distributions of the $b$-hadrons in the event sample.
One of the first things to notice is that the $p_T$ distribution has a smaller tail 
than the momentum distribution.
This is a feature that is exploited in this work.
With increasing centre-of-mass energy, the $b$-hadron production tends to be at larger pseudorapidities
and larger momenta, but the $p_T$ spectrum is less strongly affected.

Since the main features that we try to utilise in this study are related to 
the line of flight between the $b$-hadron production and decay vertices, which is denoted $\vec{F}$,
it is crucial that we model the resolution in the associated variables.
The $x$ and $y$ co-ordinates of the $b$-hadron decay vertices are smeared by $\pm 20$\micron according to a Gaussian distribution.
In the $z$ direction a larger resolution of $\pm 200$\micron is assumed.
For the production vertices we assume resolutions of $\pm 13$\micron in $x$ and $y$,
and $\pm 70$\micron in $z$.
These assumptions approximately reflect the reported peformance of the LHCb 
VELO detector~\cite{LHCb-TDR-005}.
In all subsequent studies it is required that the smeared flight length 
is larger than $3$~mm, which approximates the effect of typical
trigger and analysis selections of $b$-hadron decays by LHCb.

In order to study possible physics analysis applications, 
several $b$-hadron decays are simulated, with a
focus on \Bs mesons which are copiously produced in hadron colliders.
In order to efficiently utilise the simulated $b$-hadron samples,
all species are considered to be \Bs mesons for the purpose of studying their decays.
This is justified by the fact that the fragmentation fractions have been
measured to exhibit modest kinematic dependencies in the LHCb acceptance~\cite{LHCb-PAPER-2011-018}.
Exclusive decays of \Bs mesons to 
$D_s^-\mu^+\nu_{\mu}$ and $K^-\mu^+\nu_{\mu}$ 
are simulated with a simple phase space description, 
which is considered to be sufficiently accurate for the present study.
The $D_s^-$ mesons subsequently decay to the $K^+K^-\pi^-$ final state.
In all studies involving the \Bs decay products it is required that 
the charged final state particles satisfy $p_T > 250$~MeV, $P > 5$~GeV, $1.9 < \eta < 4.9$.
Natural units with $c=1$ are used throughout this document.
For the $K^-\mu^+\nu_{\mu}$ decay mode, tighter selection requirements are imposed. 
The muon (kaon) must satisfy $p_T > 1(0.5)$~GeV.
As a background in the study of $B_s^0 \to K^-\mu^+\nu_{\mu}$, 
we simulate the decay $B_s^0 \to (K^{*-} \to K^-\pi^0)\mu^+\nu_{\mu}$,
in which the $\pi^0$ isn't reconstructed.

%%%%%%%%%%%%%%%%%%%%%%%%%%%%%%%%%%%%%%%%%%%%%%%%%%%%%%%%%%%%%%%%%%%%%%
\section{Variables that are correlated to the b momentum}
%%%%%%%%%%%%%%%%%%%%%%%%%%%%%%%%%%%%%%%%%%%%%%%%%%%%%%%%%%%%%%%%%%%%%%

\begin{figure}[tb]\centering
\includegraphics[width=\figwidth\linewidth]{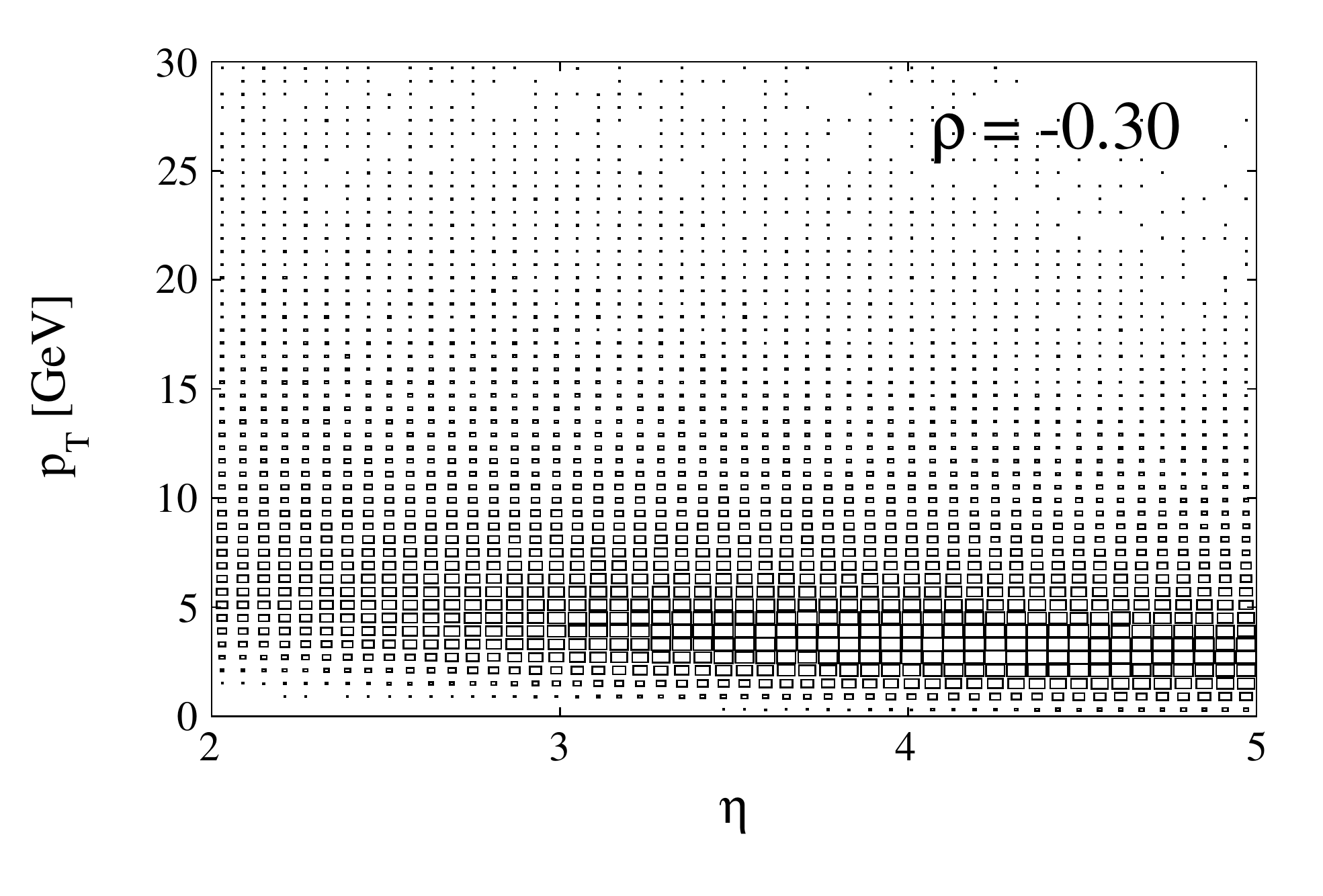}
\includegraphics[width=\figwidth\linewidth]{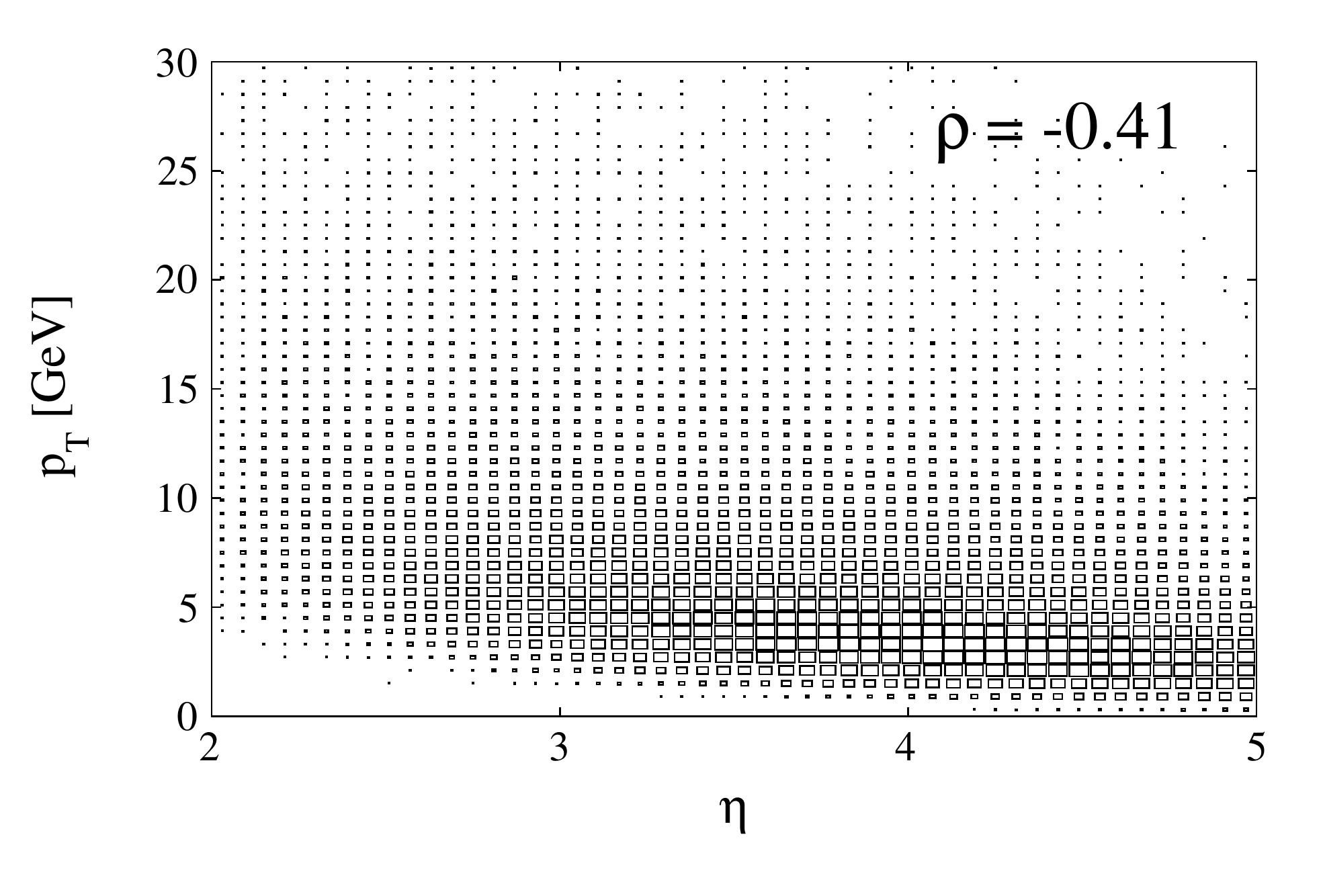}
\caption{\label{fig:B_ETA_PT}Distribution of $p_T$ versus $\eta$ 
in the simulated sample of $b$-hadrons (left) without any simulated decay
and (right) with simulated \BsToKMuNu decays that are required to satisfy
the basic selection requirements as described in the text.}
\end{figure}

\begin{figure}[tb]\centering
\includegraphics[width=\figwidth\linewidth]{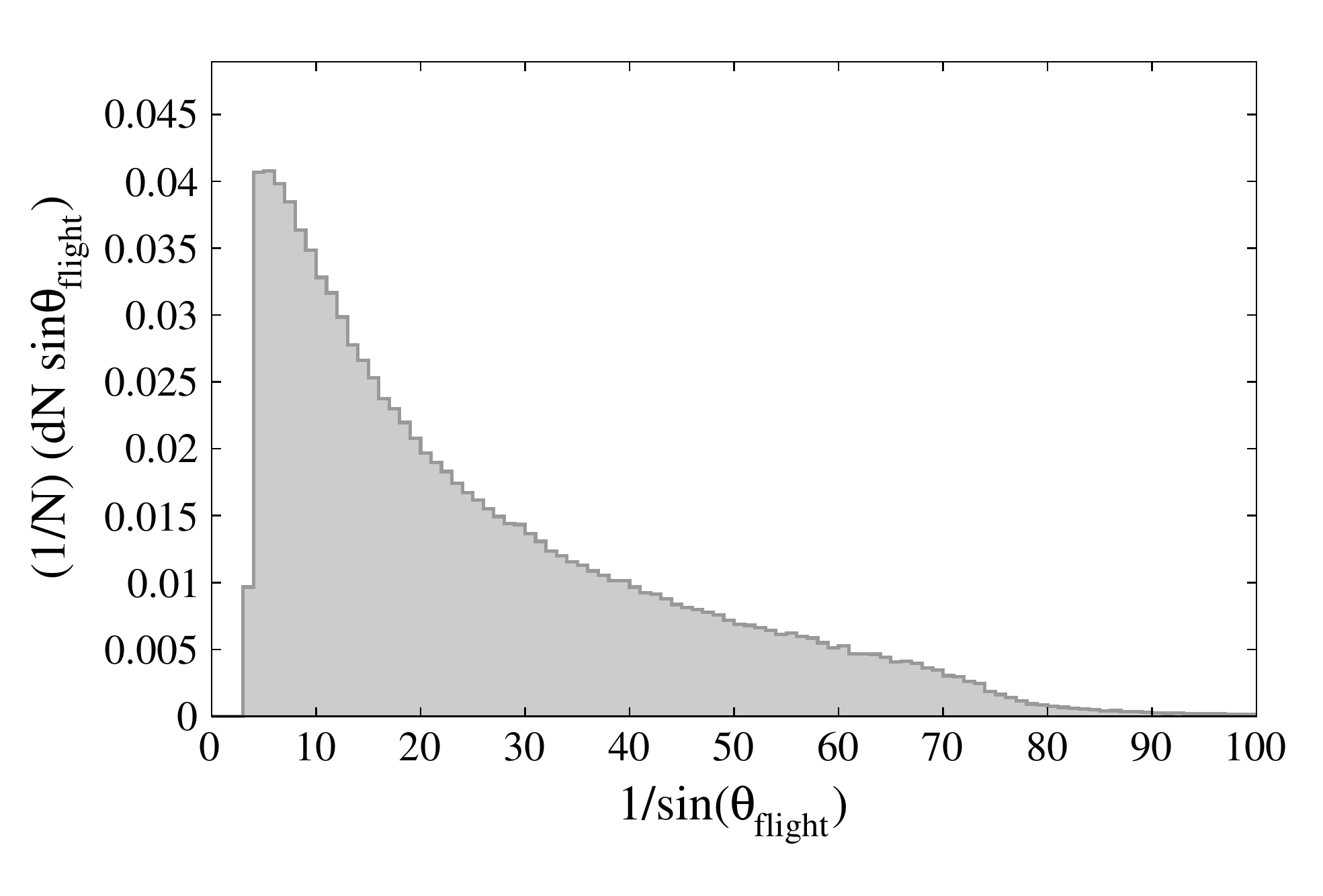}
\includegraphics[width=\figwidth\linewidth]{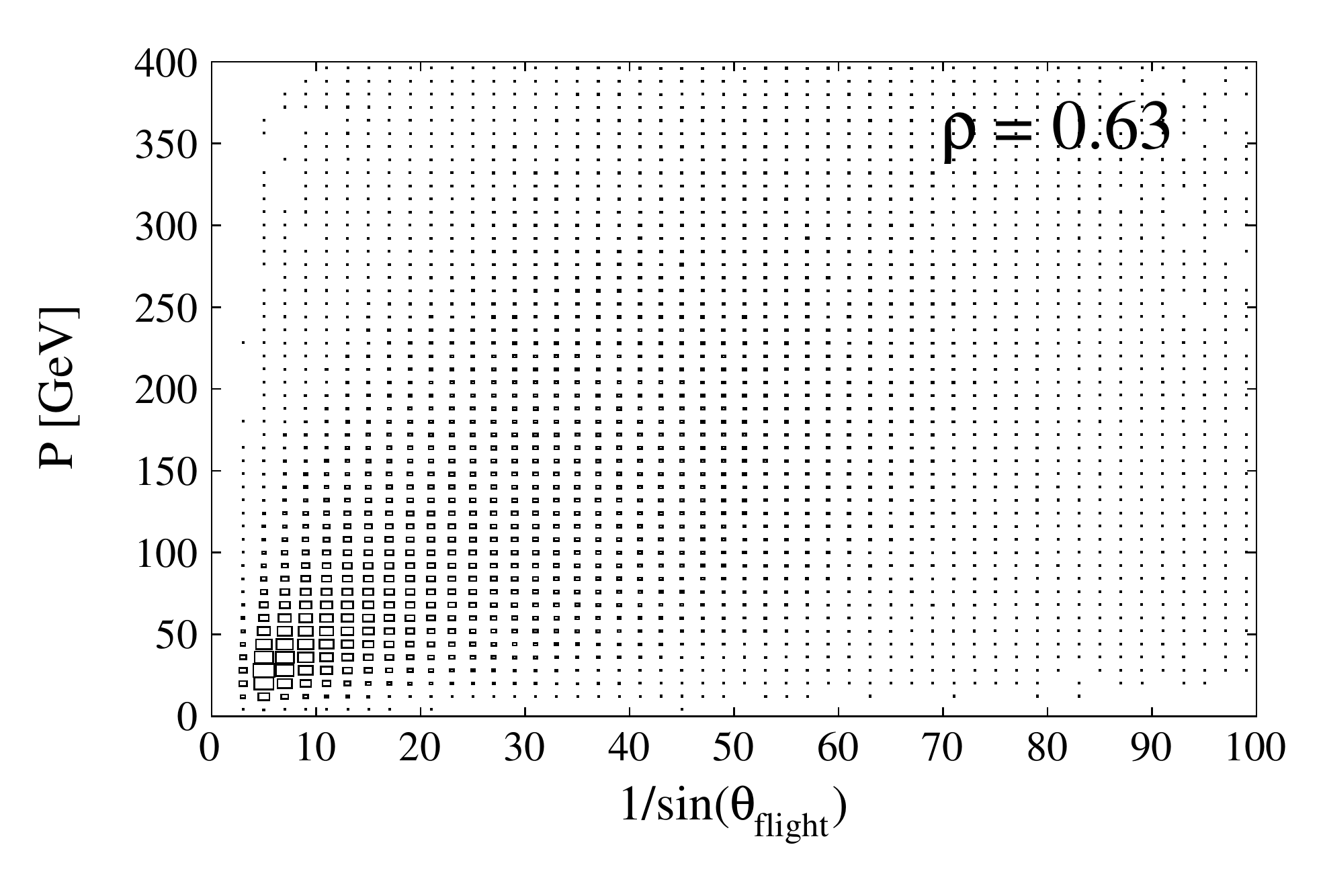}
\caption{\label{fig:REGR_INPUT_SINTHETA}The left-hand figure shows the $1/\mathrm{sin}\theta_{\rm flight}$ distribution of the simulated $b$-hadrons that are within the LHCb acceptance.
The right-hand figure  shows the distribution of the same variable versus the $b$-hadron momentum.}
\end{figure}

We attempt to identify variables that are correlated to the $b$-hadron momentum,
but strictly restrict to those that are independent of the $b$-hadron decay properties.
The single most important feature that we try to exploit is apparent in 
Fig.~\ref{fig:B_ETA_PT} (left) which shows the distribution of $p_T$ versus $\eta$.
The (anti-)correlation between the two variables is weak, with a coefficient of around 30\%,
as indicated on the figure.
It is therefore possible to estimate the momentum of the $b$-hadron as,
\begin{equation}
P = \frac{\overline{p_T}}{\sin\theta_{\rm flight}},
\end{equation}
where $\theta_{\rm flight}$ is the polar angle of the flight vector,
and it can be seen in Fig.~\ref{fig:B_NOCUTS} that $\overline{p_T} \approx 5$~GeV 
in our simulated samples.
This approximation should return a momentum estimate with a resolution function that
resembles the $p_T$ distribution in Fig.~\ref{fig:B_NOCUTS}.
In Fig.~\ref{fig:REGR_INPUT_SINTHETA} (left) the distribution of $1/\mathrm{sin}\theta_{\rm flight}$
is shown.
Fig.~\ref{fig:REGR_INPUT_SINTHETA} (right) shows that this variable has a near linear relation
to the $b$-hadron momentum with a correlation coefficient of around 65\%.
The approximation above is degraded once it is appreciated that the charged decay products
from the $b$-hadron must be within the acceptance of the detector.
Fig.~\ref{fig:B_ETA_PT} (right) shows the distribution of $p_T$ versus $\eta$
for simulated \BsToKMuNu decays that satisfy the selection cuts.
This has the effect of suppressing the region of low $p_T$ and low $\eta$,
thus increasing the magnitude of the correlation between these two variables
by around 10\%.

\begin{figure}[tb]\centering
\includegraphics[width=\figwidth\linewidth]{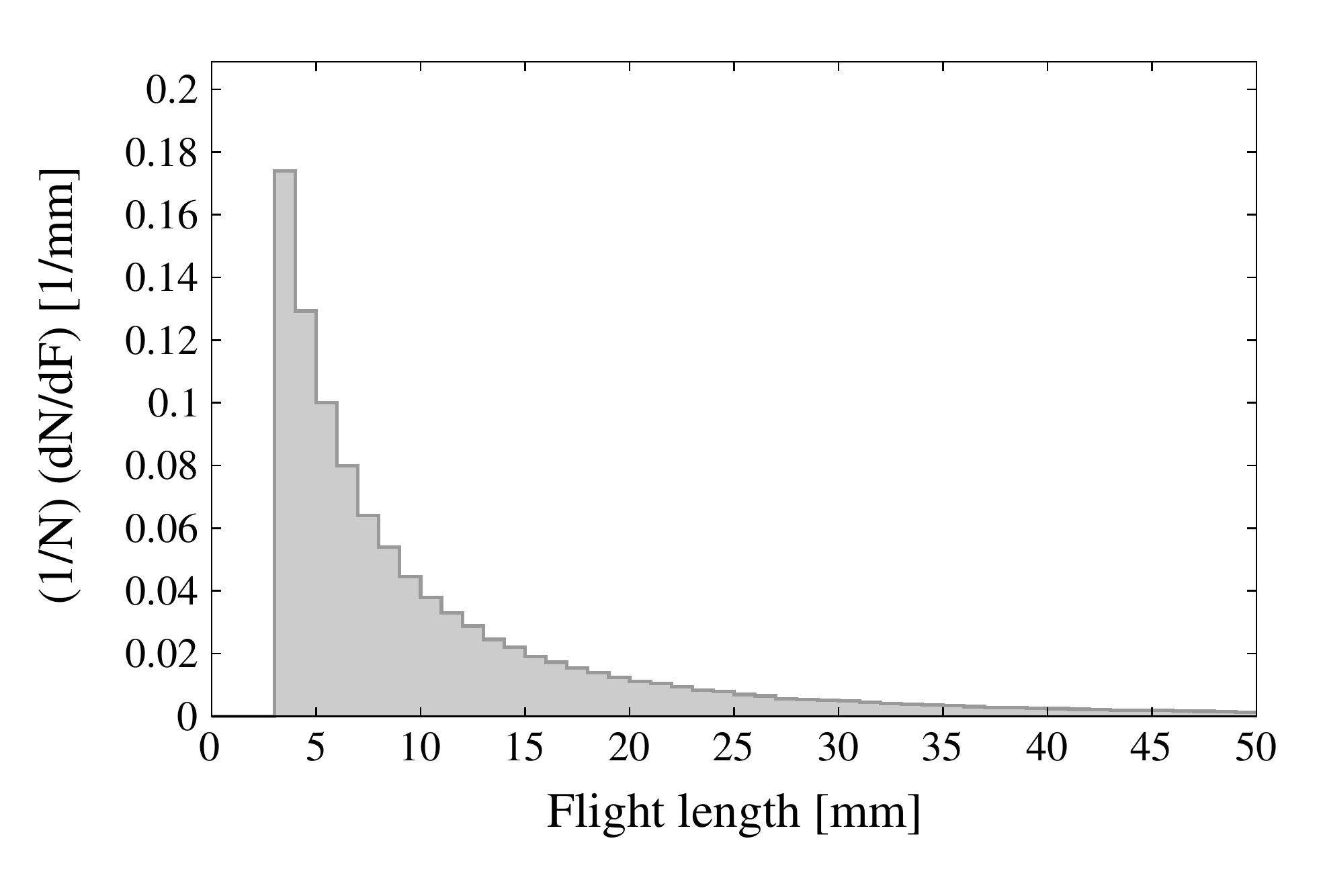}
\includegraphics[width=\figwidth\linewidth]{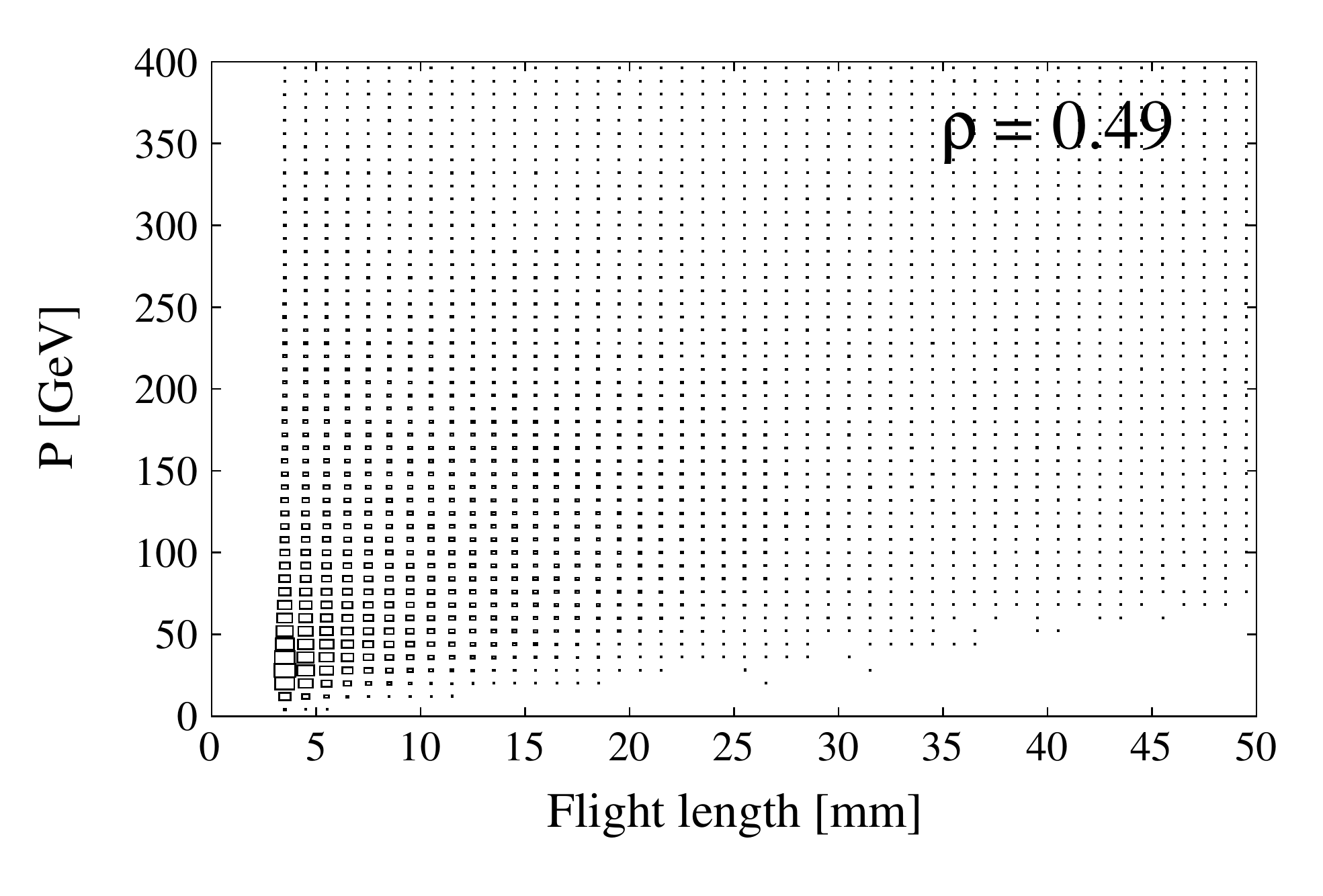}
\caption{\label{fig:REGR_INPUT_MAG}
The left-hand figure (left) shows the $|\vec{F}|$ distribution of the simulated $b$-hadrons that are within the LHCb acceptance.
The right-hand figure (right) shows the distribution of the same variable versus the $b$-hadron momentum.}
\end{figure}

The flight length, $|\vec{F}|$, of a $b$-hadron of mass $M$ and decay time $t$ can be directly related to the momentum according to,
\begin{equation}
P = \frac{M |\vec{F}|}{t}.
\end{equation}
Fig.~\ref{fig:REGR_INPUT_MAG} (left) shows the distribution of $|\vec{F}|$,
in which our requirement of at least $3$~mm is clearly visible.
Fig.~\ref{fig:REGR_INPUT_MAG} (right) shows that this variable is correlated with the
momentum with a coefficient of around 50\%.

We consider the use of information from other reconstructed particles in the event.
It is obvious that in the hypothetical case of a detector with 4$\pi$ angular coverage
and perfect efficiency and resolution, the $b$-hadron $p_T$ could be inferred from the  
transverse momentum balance.
Considering the LHCb detector, and the most optimistic use of all kinematic information
from the reconstructible particles, we can only achieve a correlation of around 20\% between
the missing $p_T$ and the $p_T$ of the signal $b$.
As an alternative, we consider the possibility to reconstruct the $\bar{b}$-hadron that is produced in association with the signal $b$.
Even at $b(\bar{b})$-quark level the naive $p_T$ balance between the 
$b$ and $\bar{b}$ is spoilt by the broad $b\bar{b}$ $p_T$ spectrum.
Various combinations of reconstructing the signal $b$ and associated 
$\bar{b}$ at hadron or jet level are considered.
%\footnote{In the case of the signal $b$-hadron at jet level, we consider particles within a hollow cone around the $b$,
%in order to avoid picking up the $b$-hadron decay products.}
Even before considering the inefficiency of reconstructing the associated $\bar{b}$
this approach does not seem promising.

We are left with the conclusion that there are only two pieces of information
related to the $b$-hadron flight vector, namely $1/\sin\theta_{\rm flight}$ and $|\vec{F}|$, 
which are of value in an estimator of the $b$-hadron momentum.
In the following section we utilise them in a regression algorithm.

%%%%%%%%%%%%%%%%%%%%%%%%%%%%%%%%%%%%%%%%%%%%%%%%%%%%%%%%%%%%%%%%%%%%%%
\section{Multivariate regression analysis}
\label{sec:Regr}
%%%%%%%%%%%%%%%%%%%%%%%%%%%%%%%%%%%%%%%%%%%%%%%%%%%%%%%%%%%%%%%%%%%%%%

\begin{figure}[tb]\centering
\includegraphics[width=\figwidth\linewidth]{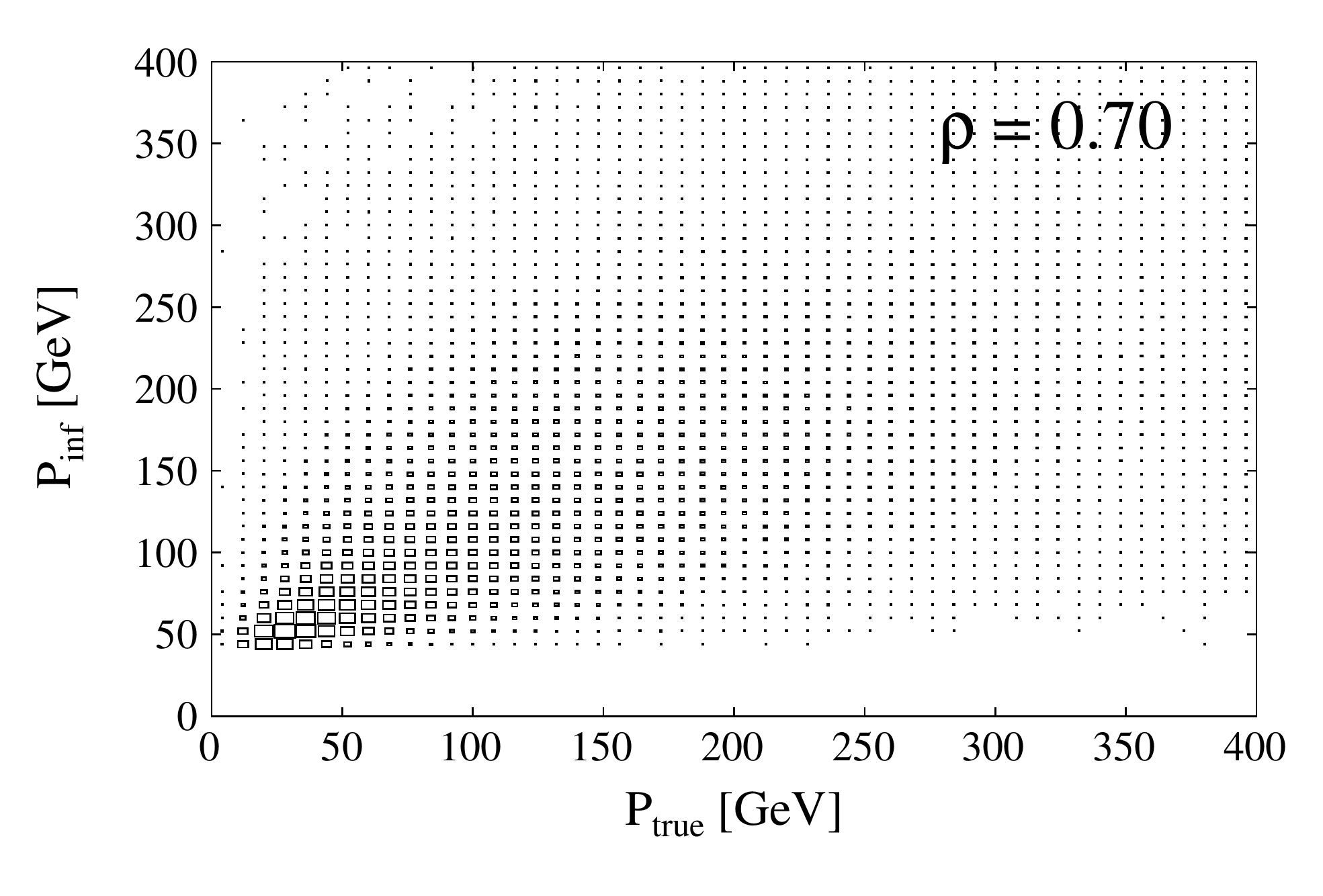}
\caption{\label{fig:REGR_PLOTS_2D}The distribution of \Pinf versus the true $b$-hadron momentum.}
\end{figure}

The two flight variables described in the previous section, $1/\sin\theta_{\rm flight}$ and $|\vec{F}|$, 
are considered in a multivariate regression analysis in order to infer the momenta of the $b$-hadrons.
A simple least squares linear regression algorithm, as implemented in the {\texttt sklearn} package~\cite{scikit-learn}, is used.
This algorithm is trained on a randomly selected subset of the simulated event sample.
The independent data are used to evaluate the performance of the algorithm in estimating the $b$-hadron momentum
from the values of the two flight variables.
Fig.~\ref{fig:REGR_PLOTS_2D} shows the distribution of the inferred $b$-hadron momentum, \Pinf, 
versus the true $b$-hadron momentum.
The correlation coefficient is around 70\%.
In Tab.~\ref{tab:Correlations}, the correlation coefficients 
between $P_{\rm true}$ and the two flight variables are listed for the three centre of 
mass energies and various selection requirements on the simulated \BsToKMuNu decays.
Also listed are the correlations between $P_{\rm true}$ and the inferred momentum
that would be returned by the regression using only $1/\sin\theta_{\rm flight}$, 
which is denoted \PinfTheta.
It can be seen that as expected these values are close to the corresponding correlations with the raw flight angle variable itself.
The final column of Tab.~\ref{tab:Correlations} lists the correlations between $P_{\rm true}$ and \Pinf.
It can be seen that the combination of the two variables in the regression algorithm
increases the correlation by around 10\% compared to the more powerful angular variable alone.
Hardly any dependence on the centre-of-mass energy is seen.
There is a degradation of the correlations of up to 10\% when applying 
the acceptance and selection requirements on the charged decay products of the simulated \BsToKMuNu decays.

\begin{figure}[tb]\centering
\includegraphics[width=\figwidth\linewidth]{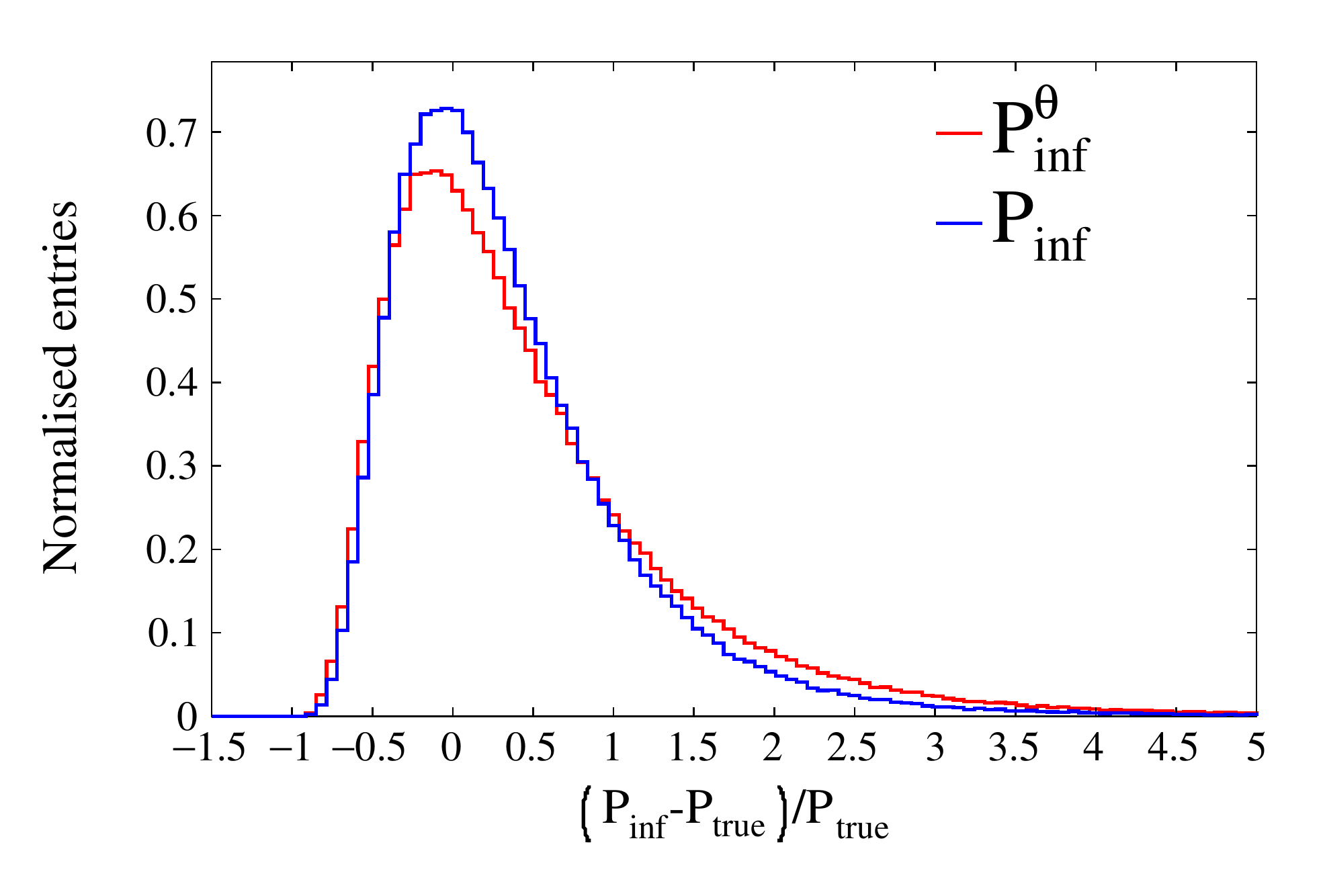}
\includegraphics[width=\figwidth\linewidth]{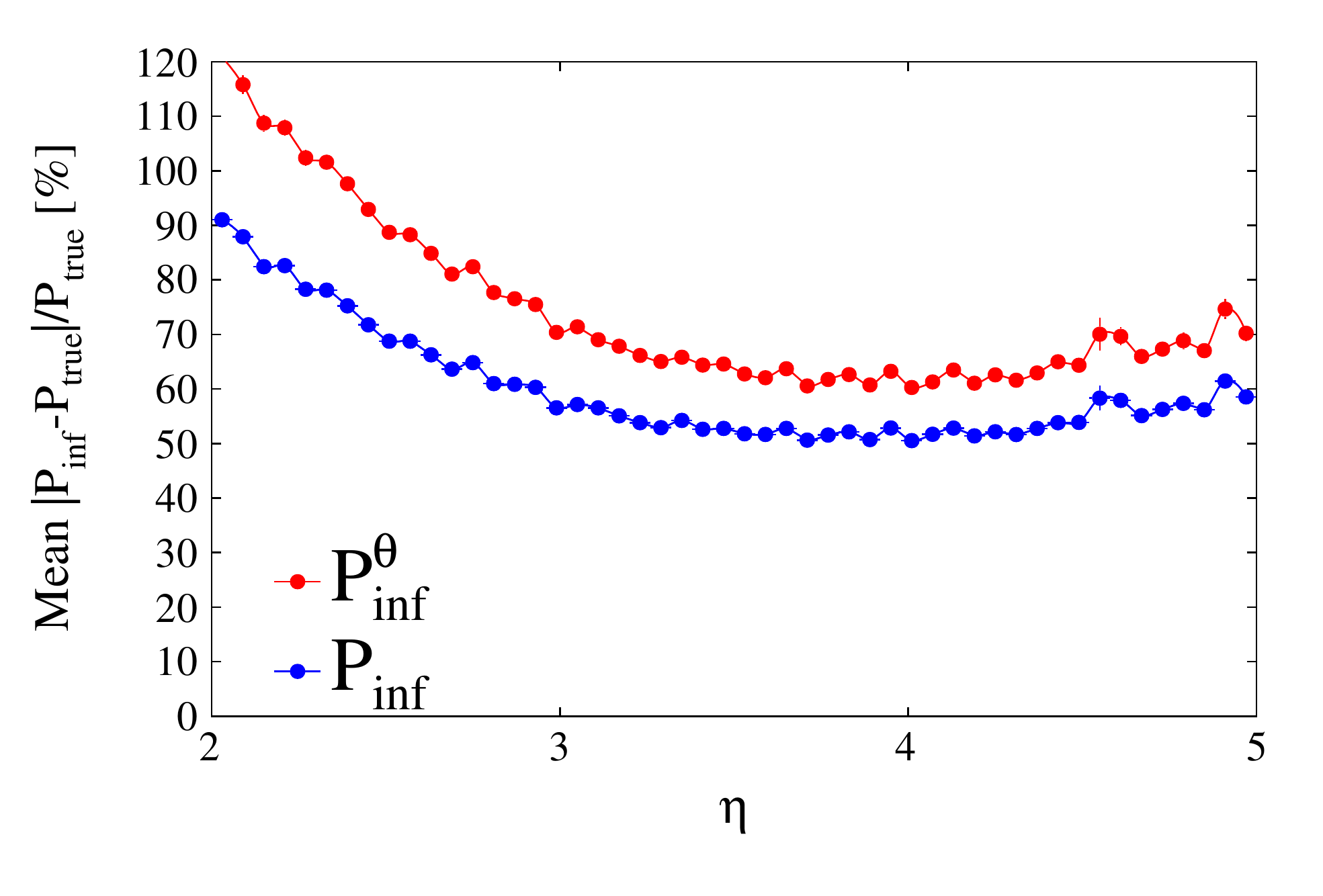}
\caption{\label{fig:REGR_PLOTS_Res}The left-hand figure shows the distribution of $(P_{\rm inf}-P_{\rm true})/P_{\rm true}$. The right-hand figure shows how the profile of $|P_{\rm inf}-P_{\rm true}|/P_{\rm true}$ varies with $\eta$.
Both figures include the corresponding entries for \PinfTheta.}
\end{figure}

%The inclusion of $|\vec{F}|$ in the two-variable version
%increases the correlation between \Pinf and $P_{\rm true}$ by a few percent.
Fig.~\ref{fig:REGR_PLOTS_Res} (left) shows the distribution of $(P_{\rm inf}-P_{\rm true})/P_{\rm true}$
and the corresponding distribution for \PinfTheta instead of \Pinf.
As expected the shapes of these distributions roughly resemble the underlying $b$-hadron $p_T$ spectrum
shown in Fig.~\ref{fig:B_NOCUTS}.
In Fig.~\ref{fig:REGR_PLOTS_Res} (right) the corresponding profiles of the mean 
$|P_{\rm inf}-P_{\rm true}|/P_{\rm true}$ are shown as a function of $\eta$.
The resolution of \Pinf is around 60\% and exhibits some dependence on $\eta$.
It is about 10--20\% improved compared to that of \PinfTheta which neglects
the decay length information.

%We further check the robustness of the method with respect to our assumptions 
%on the vertex resolution.
%Even for variations of up to three orders of magnitude there is a negligible effect on the performance.

\begin{table}\centering
\caption{\label{tab:Correlations}The coefficients of correlation 
between the true $b$-hadron momentum, and the raw flight variables and the inferred momentum
from the regression. For each centre-of-mass energy, as indicated in the first column,
the first row corresponds to only the basic flight length and acceptance requirements
on the $b$-hadron. The second and third rows sequentially apply $P,p_T$ and $\eta$
requirements on the charged decay products in the simulated \BsToKMuNu decays.}
\begin{tabular}{rlcccc}
\hline
\hline
& & \multicolumn{4}{c}{Correlation coefficient}\\
$\sqrt{s}$ & Cuts& $1/\sin\theta$& $|F|$& $P_{\rm inf}^{\theta}$& $P_{\rm inf}$\\
\hline
\hline
7 TeV & None& 0.63& 0.50& 0.61& 0.69\\
7 TeV & $P,p_T$& 0.59& 0.52& 0.58& 0.69\\
7 TeV & $P,p_T,\eta$& 0.52& 0.50& 0.51& 0.65\\
\hline
13 TeV & None& 0.63& 0.49& 0.63& 0.70\\
13 TeV & $P,p_T$& 0.60& 0.50& 0.59& 0.69\\
13 TeV & $P,p_T,\eta$& 0.53& 0.48& 0.53& 0.65\\
\hline
100 TeV & None& 0.62& 0.48& 0.63& 0.69\\
100 TeV & $P,p_T$& 0.59& 0.50& 0.60& 0.69\\
100 TeV & $P,p_T,\eta$& 0.53& 0.48& 0.54& 0.65\\
\hline
\hline
\end{tabular}

\end{table}

%%%%%%%%%%%%%%%%%%%%%%%%%%%%%%%%%%%%%%%%%%%%%%%%%%%%%%%%%%%%%%%%%%%%%%
\section{Physics applications}
%%%%%%%%%%%%%%%%%%%%%%%%%%%%%%%%%%%%%%%%%%%%%%%%%%%%%%%%%%%%%%%%%%%%%%

In this section, several physics applications are considered.
Sect.~\ref{sec:Decays} describes an application to the study 
the decay \BsToKMuNu.
The $b$-hadron momentum estimate is used to resolve the quadratic ambiguity and enhance the resolution
in the kinematic quantities describing the $b$-hadron decay.
Sect.~\ref{subsec:discrimination} describes an attempt to use the momentum estimate 
directly to define variables that distinguish between different classes of decays.
As an example, we consider separating \BsToKMuNu from \BsToKstMuNu.
Sect.~\ref{subsec:mixing} reports on the use of the regression based choice of quadratic solution for the
$b$-hadron momentum to make differential measurements of the asymmetry between
the production rates of \Bs versus \Bsb mesons, which requires resolution of \BsBsb oscillations.

\subsection{Application to the study of semileptonic b decays}
\label{sec:Decays}

As mentioned in the introduction, the kinematic properties of a decay with
a single unreconstructed particle of known mass can be solved up to a quadratic 
ambiguity by imposing momentum balance against the visible system with respect to the flight vector,
and assuming the mass of the $b$~\cite{Dambach:2006ha}.
The vector sum of the momenta of the reconstructed decay products is referred to 
as the visible momentum, and is denoted \vecPVis.
The corresponding missing momentum associated with the unreconstructed decay products
is denoted \vecPMiss.
The visible momentum is decomposed into components transverse to
and along the flight vector,
\begin{align}
\PTVis &= \left|\vecPVis \times \frac{\vec{F}}{|\vec{F}|}\right|, \\
\PLVis &= \vecPVis \cdot \frac{\vec{F}}{|\vec{F}|}.
\end{align}
The transverse component of the missing momentum, \PTMiss, 
is fixed to be equal to \PTVis.
Assuming that the $b$-hadron has a mass $m$, and that there is a single massless unreconstructed particle
in the final state, one can derive a quadratic equation in \PLMiss,
\begin{align}
\PLMiss &= -a \pm \sqrt{r},
\end{align}
with, 
\begin{align}
a &= \frac{ \PLVis \left(m^2 - \MVis^2 - 2(\PTVis)^2\right)}{2\left((\PLVis)^2 - \EVis^2\right)},\\
r &= \frac{ \EVis^2 \left( m^2 - \MVis^2 - 2(\PTVis)^2 \right)^2 }{ 4 \left( (\PLVis)^2 - \EVis^2 \right)^2 } + \frac{ \left( \EVis\PTVis \right)^2 }{ (\PLVis)^2 - \EVis^2},
\end{align}
where \EVis and \MVis are the visible energy and mass, respectively.
This yields two solutions for the $b$-hadron momentum,
\begin{align}
\label{eq:PpPm}
P_{+} &= \PLVis -a + \sqrt{r},\\
P_{-} &= \PLVis -a - \sqrt{r}.
\end{align}
The vertex resolution renders a fraction of decays with nonphysical negative values of $r$.
These are excluded in the following analysis.

\begin{figure}[tb]\centering
\includegraphics[width=\widefigwidth\linewidth]{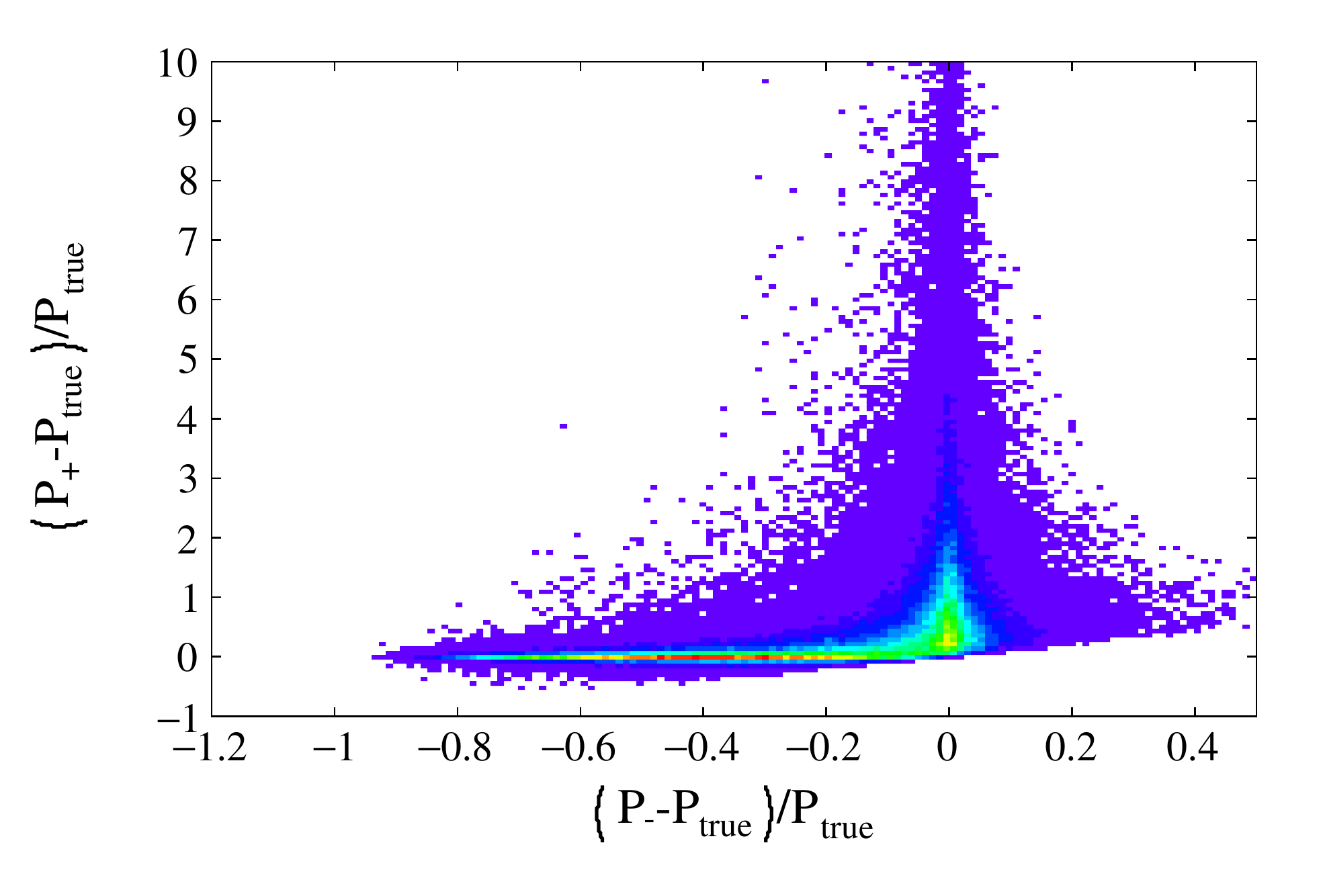}
\caption{\label{fig:PAPB}The distribution of $(P_{+}-P_{\rm true})/P_{\rm true}$ versus $(P_{-}-P_{\rm true})/P_{\rm true}$ in the subset of simulated \BsToKMuNu decays that satisfy the selection requirements as described in the text.}
\end{figure}

We now consider \BsToKMuNu decays with the already described selection requirements.
Fig.~\ref{fig:PAPB} shows the distribution of $(P_{+}-P_{\rm true})/P_{\rm true}$
versus $(P_{-}-P_{\rm true})/P_{\rm true}$.
A horizontal (vertical) band can be clearly seen for the cases in which $P_{+}$ ($P_{-}$)
is the correct solution.
The effect of the vertex smearing is clearly visible but the two bands are nevertheless well separated.
One can see that an independent estimate of the momentum can help to select the correct solution
even if it has a modest resolution.

\begin{figure}[tb]\centering
\includegraphics[width=\figwidth\linewidth]{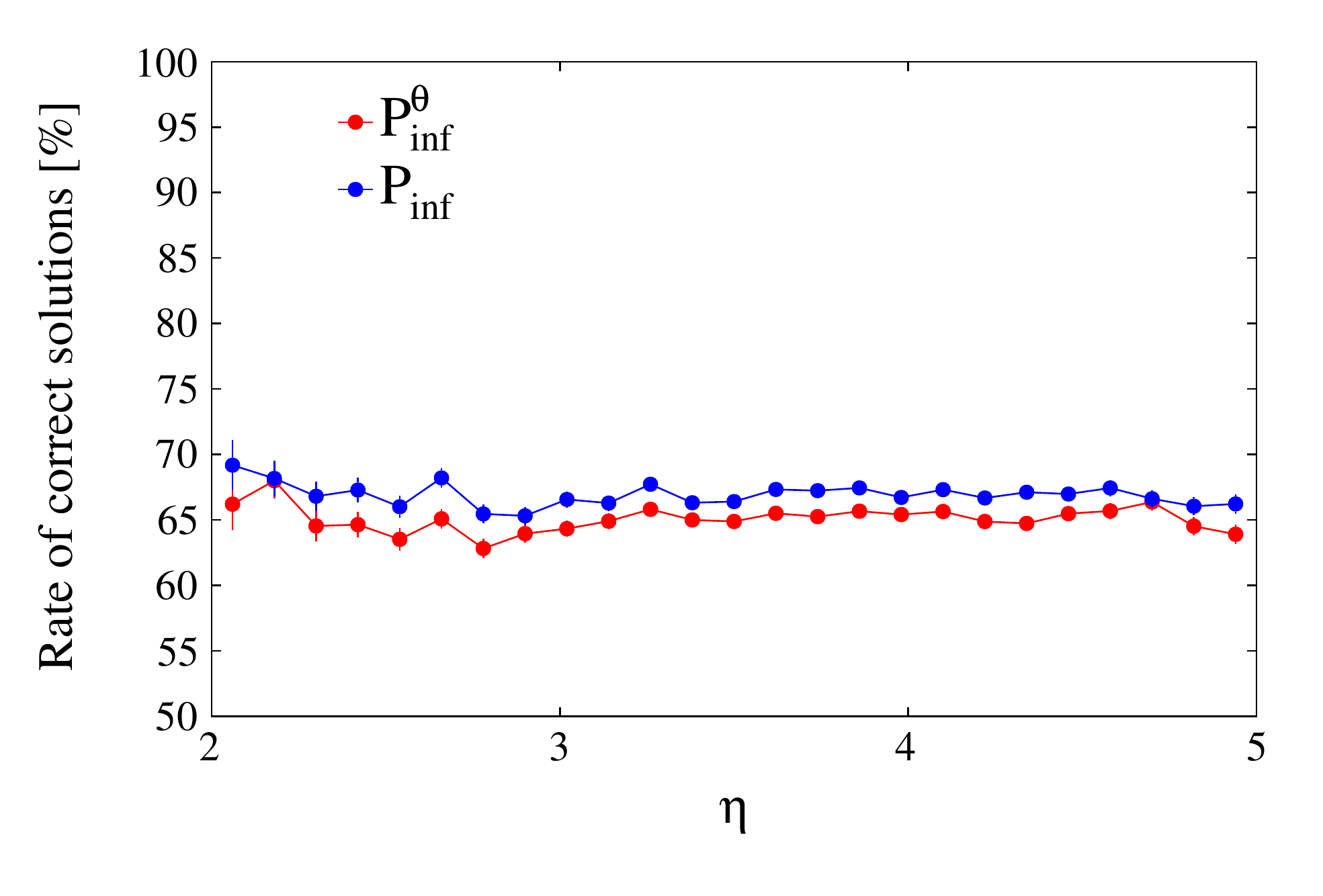}
\includegraphics[width=\figwidth\linewidth]{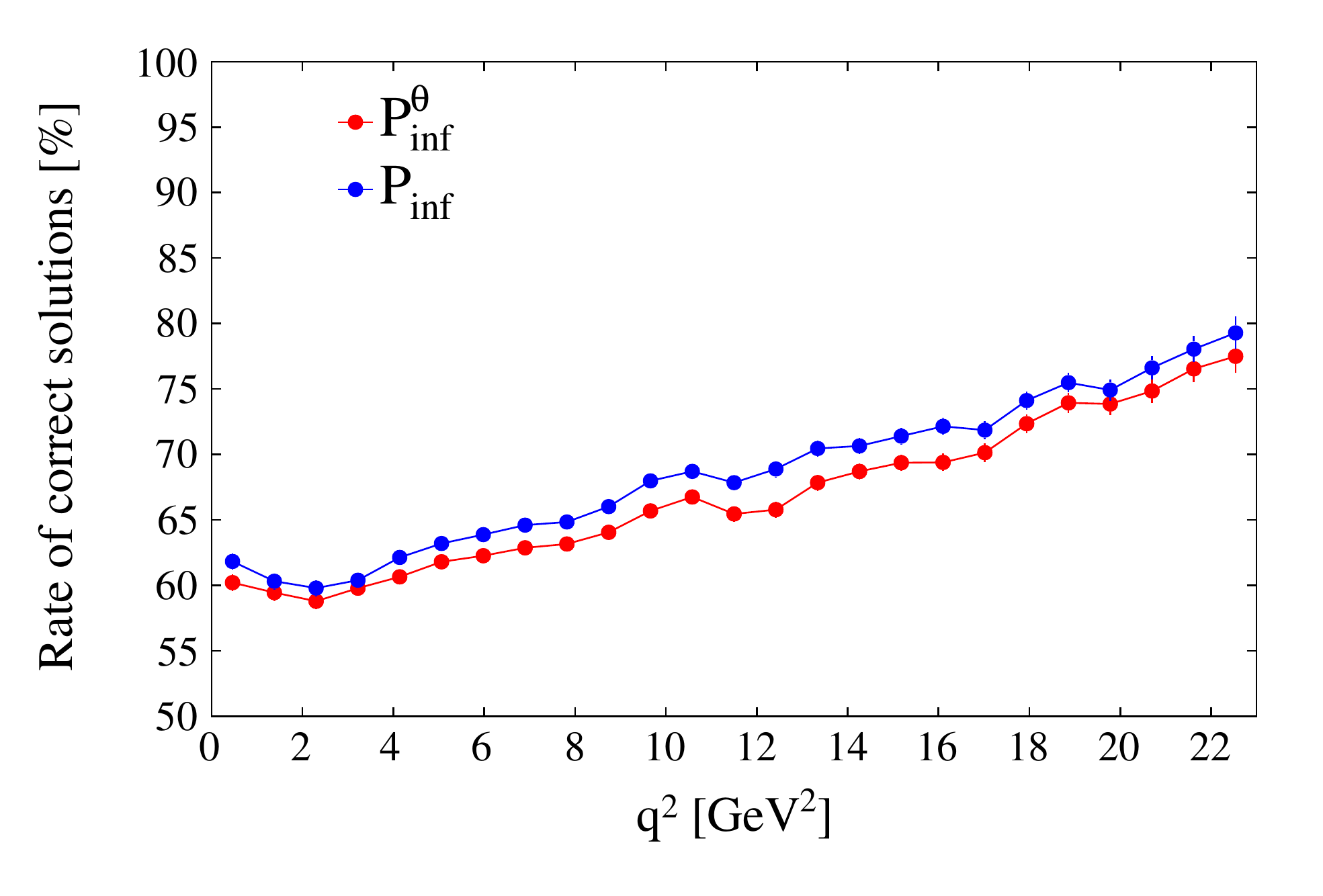}
\caption{\label{fig:Q2resProf}The rate at which the correct $b$-hadron momentum solution is chosen,
as a function of various kinematic properties of the $b$.}
\end{figure}

\begin{figure}[tb]\centering
\includegraphics[width=\figwidth\linewidth]{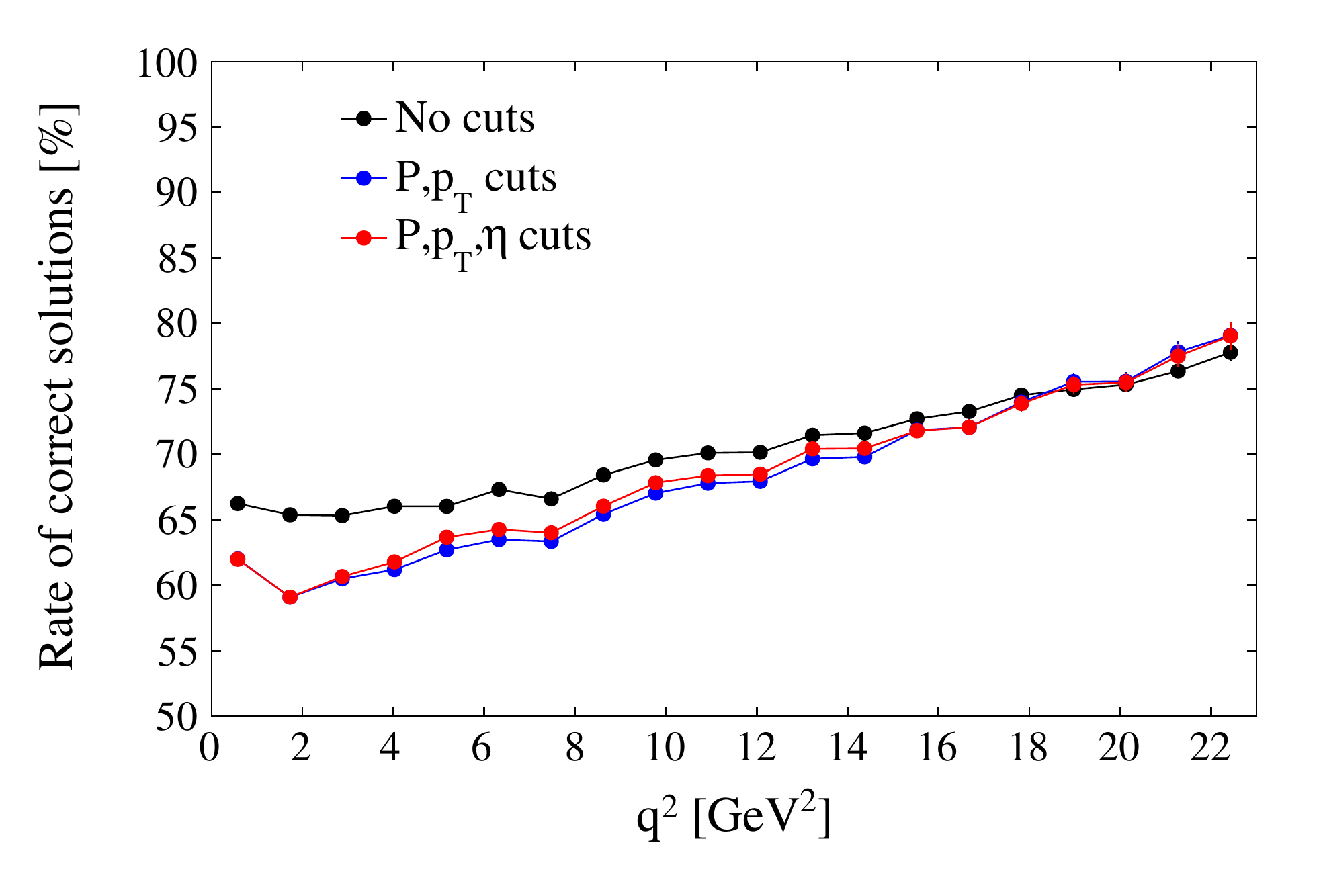}
\caption{\label{fig:Q2resProfDiffCuts}The rate at which the correct $b$-hadron momentum solution is chosen,
as a function of $q^2$. 
Separate points are show for only the $b$-hadron level selection cuts,
and for sequential application of $P, p_T$ and $\eta$ cuts on the charged final state
particles from the simulated \BsToKMuNu decays.}
\end{figure}

We can now test how often our regression based estimate of the $b$-hadron momentum 
is closer to the correct solution.
Fig.~\ref{fig:Q2resProf} shows the rate of correct choices
as a function of $\eta$ and $q^2$.
The average rate of correct solutions is around 70\%.
Fig.~\ref{fig:Q2resProfDiffCuts} shows that despite the apparent reduction 
in correlation between the true and inferred momentum when applying acceptance
cuts (see Tab.~\ref{tab:Correlations}), the rate of correct solutions is not strongly affected.
It can also be seen that the rate of correct solutions exhibits a dependence on
$q^2$ since the two solutions become more distinct at higher $q^2$.
There is a change of roughly 15\% in the absolute rate over the full $q^2$ range.

\begin{figure}[tb]\centering
\includegraphics[width=\figwidth\linewidth]{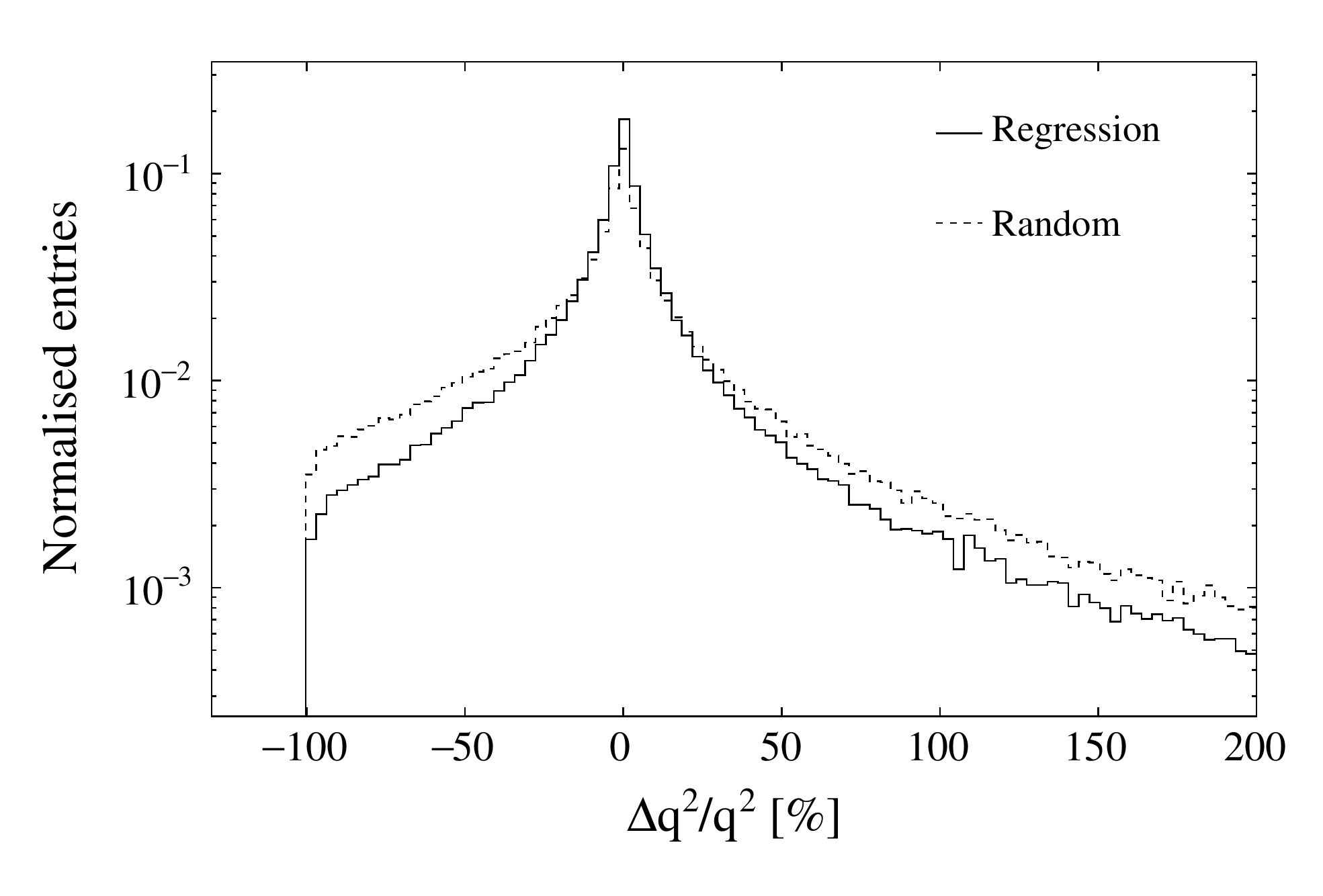}
\caption{\label{fig:Q2res}A comparison of the $q^2$ resolution achieved with 
the regression based method versus a random choice of quadratic solutions.}
\end{figure}

\begin{figure}[tb]\centering
\includegraphics[width=\figwidth\linewidth]{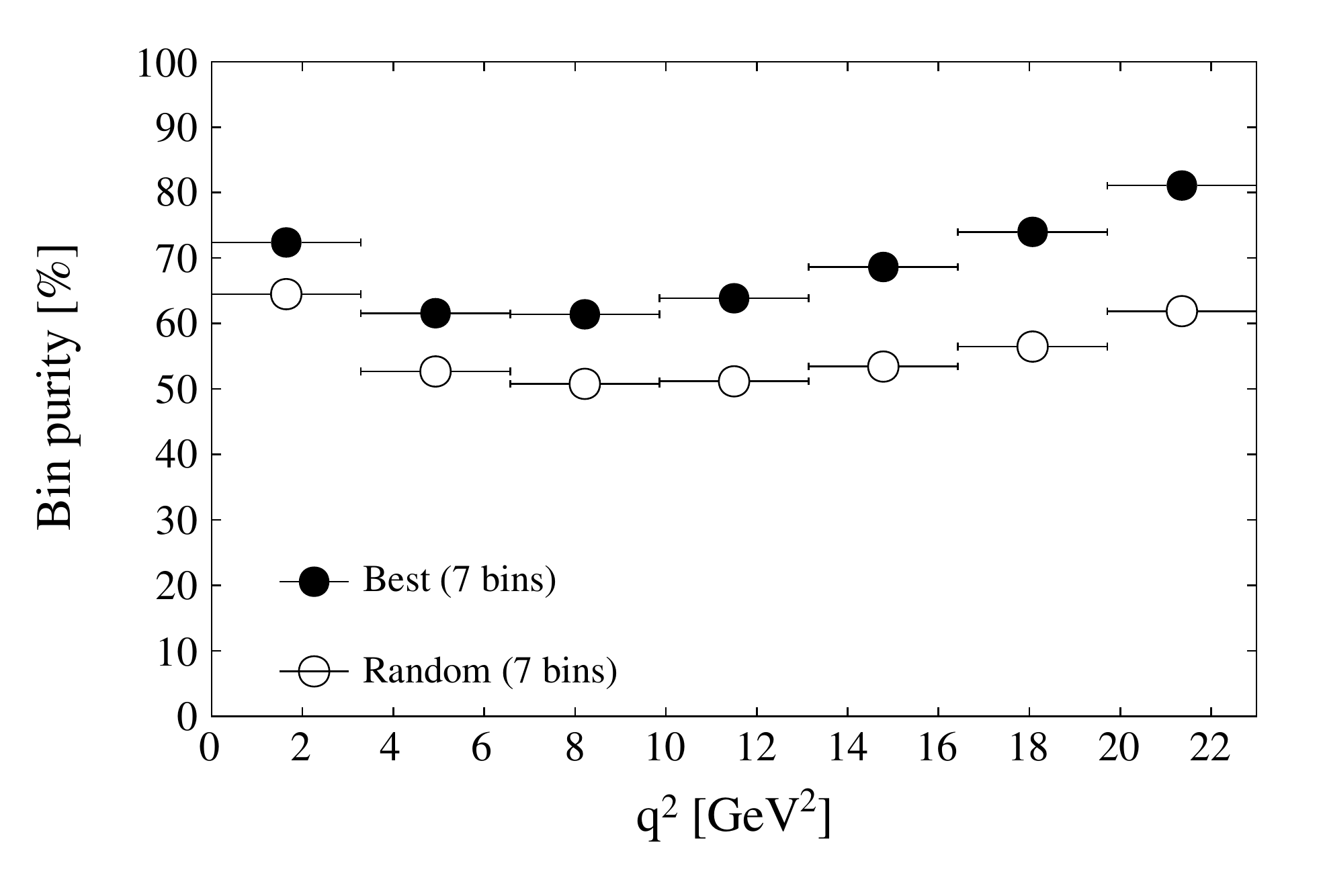}
\includegraphics[width=\figwidth\linewidth]{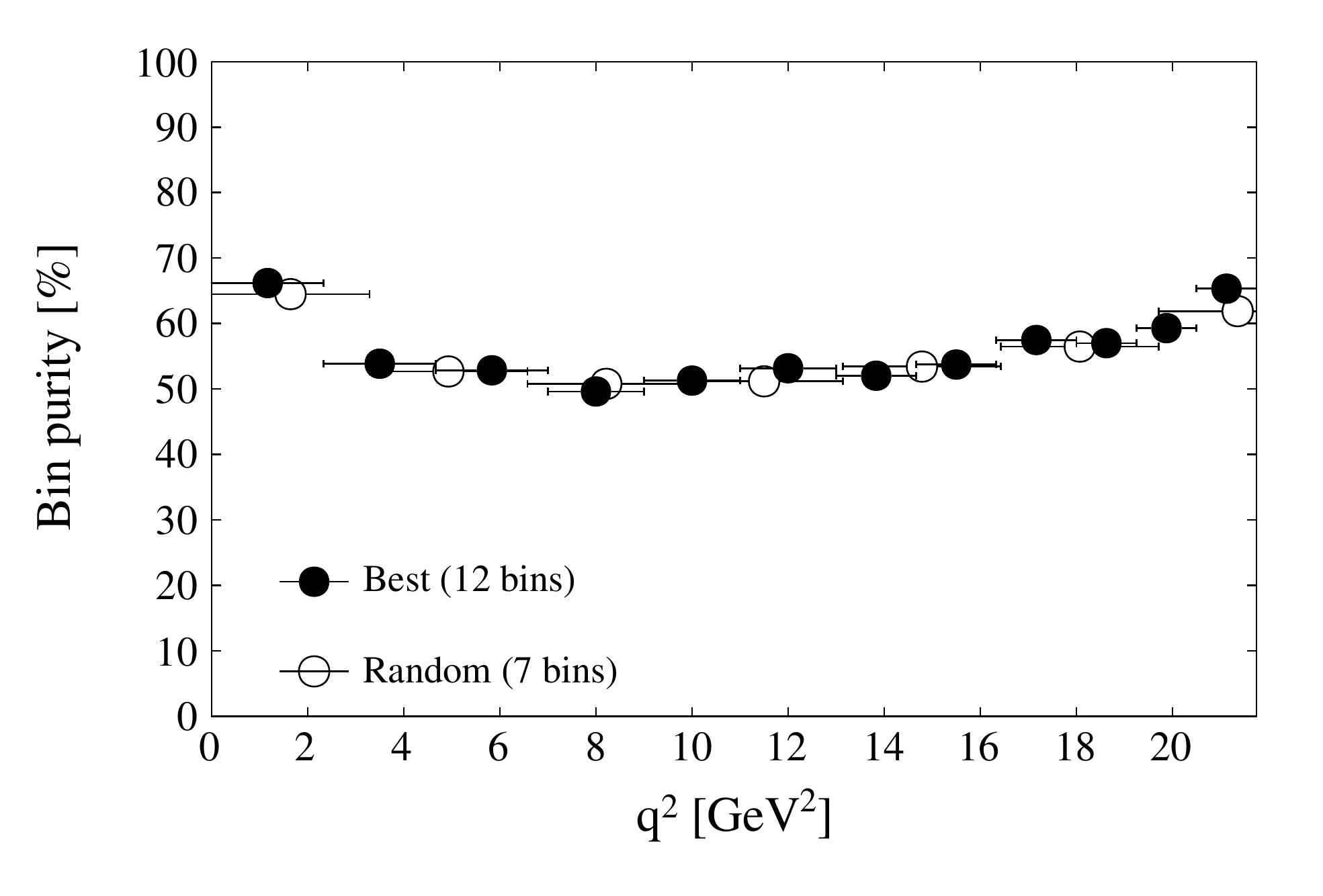}
\caption{\label{fig:Q2purity}The bin purity, as defined in the text, as a function of the true $q^2$.
The open markers correspond to a random choice of the two quadratic solutions
whereas the closed markers correspond to the regression based analysis.
Left: seven bins in both cases.
Right: twelve bins in the regression case.}
\end{figure}

It would be desirable to measure the differential decay rate as a function of $q^2$,
but this requires unfolding for the finite $q^2$ resolution,
which will inevitably introduce associated uncertainty.
Fig.~\ref{fig:Q2res} compares the $q^2$ resolution 
that is obtained with a random choice of solutions
versus a choice based on \Pinf.
A useful figure of merit in unfolding problems is the bin purity.
For a given bin in the true quantity, we define its purity as the fraction of entries for which the 
reconstructed quantity also falls into the same bin.
Fig.~\ref{fig:Q2purity} compares the $q^2$ bin purities for the random quadratic solution 
versus the best solution with our method.
In Fig.~\ref{fig:Q2purity} (left) seven equal width bins are used over the full $q^2$ range,
and it can be seen that our method achieves a 10-20\% increase in purity.
Fig.~\ref{fig:Q2purity} (right) shows that twelve appropriately defined bins
could yield the same purity as for seven bins with the random approach.
Particularly narrow bins can be used in the high $q^2$ region.

\subsection{Discrimination between different classes of semileptonic decays}
\label{subsec:discrimination}
\begin{figure}[tb]\centering
\includegraphics[width=\figwidth\linewidth]{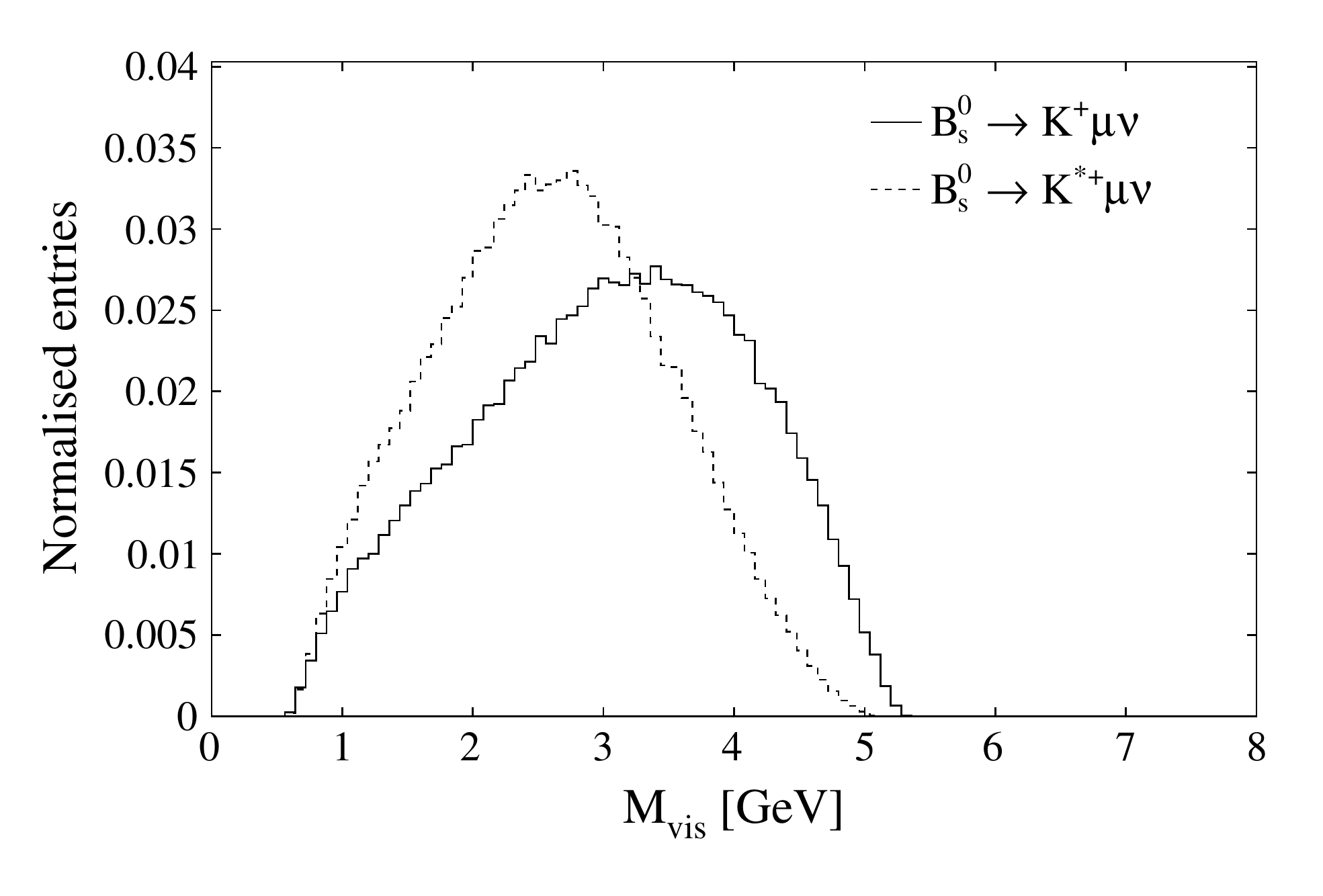}
\includegraphics[width=\figwidth\linewidth]{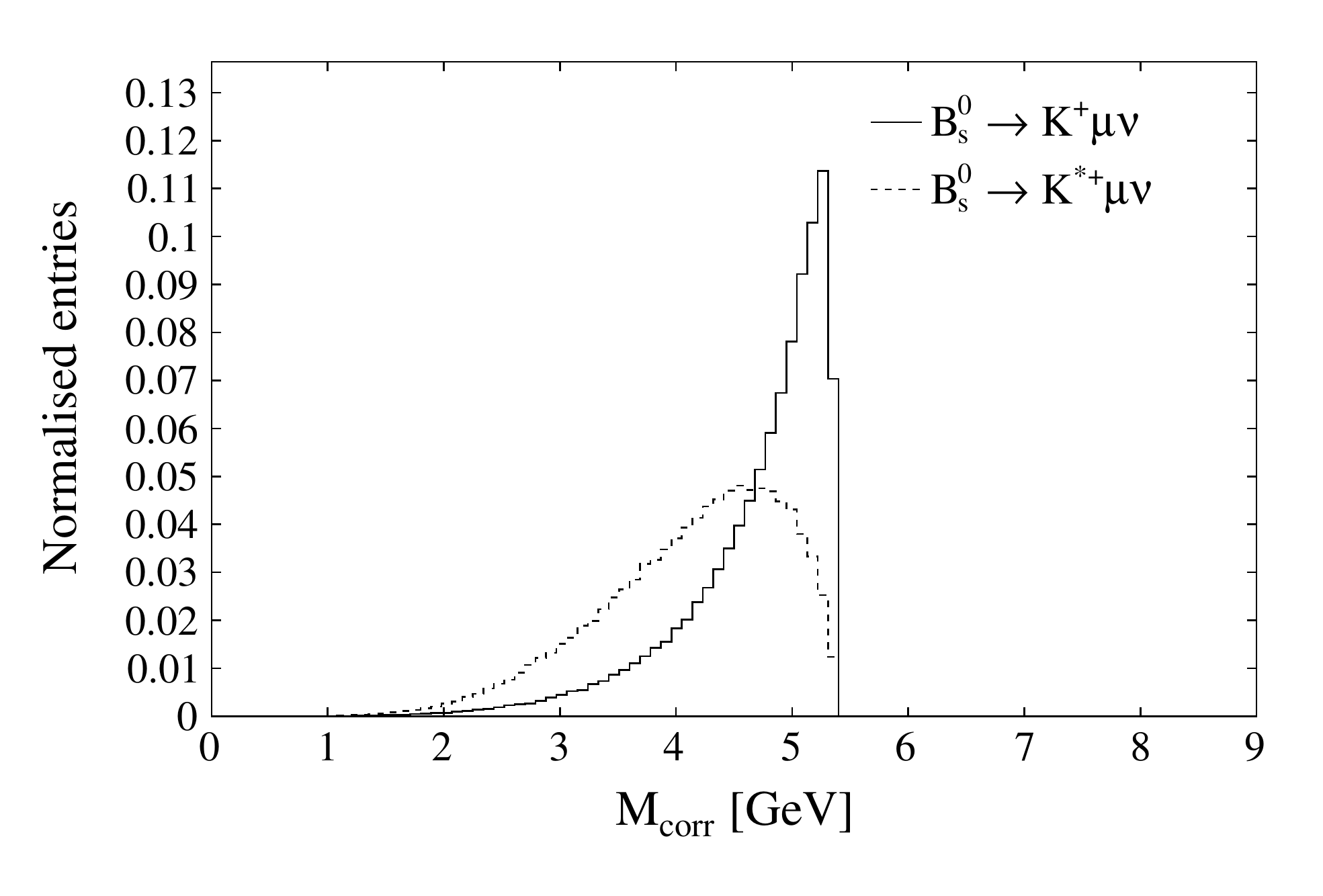}
\includegraphics[width=\figwidth\linewidth]{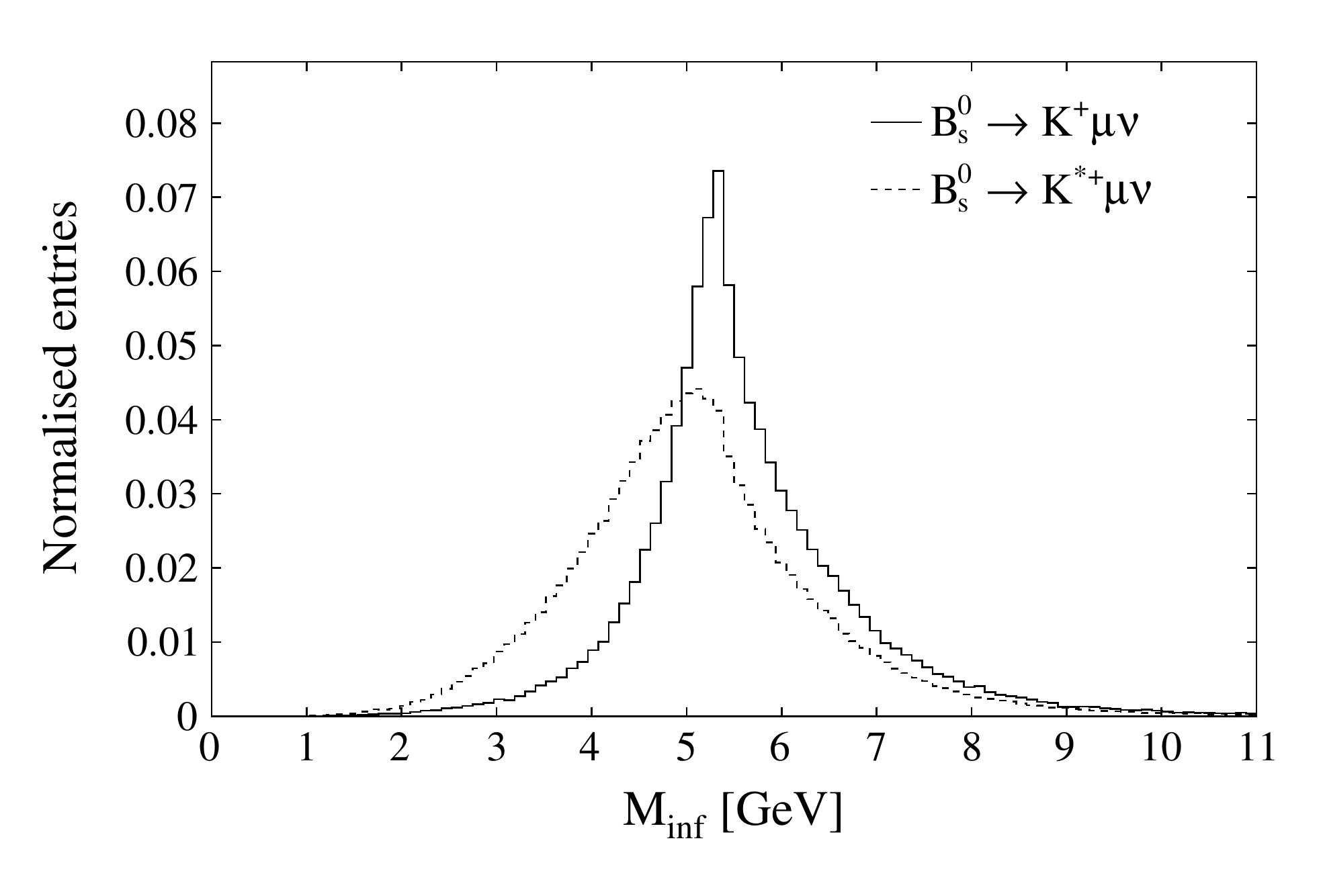}
\includegraphics[width=\figwidth\linewidth]{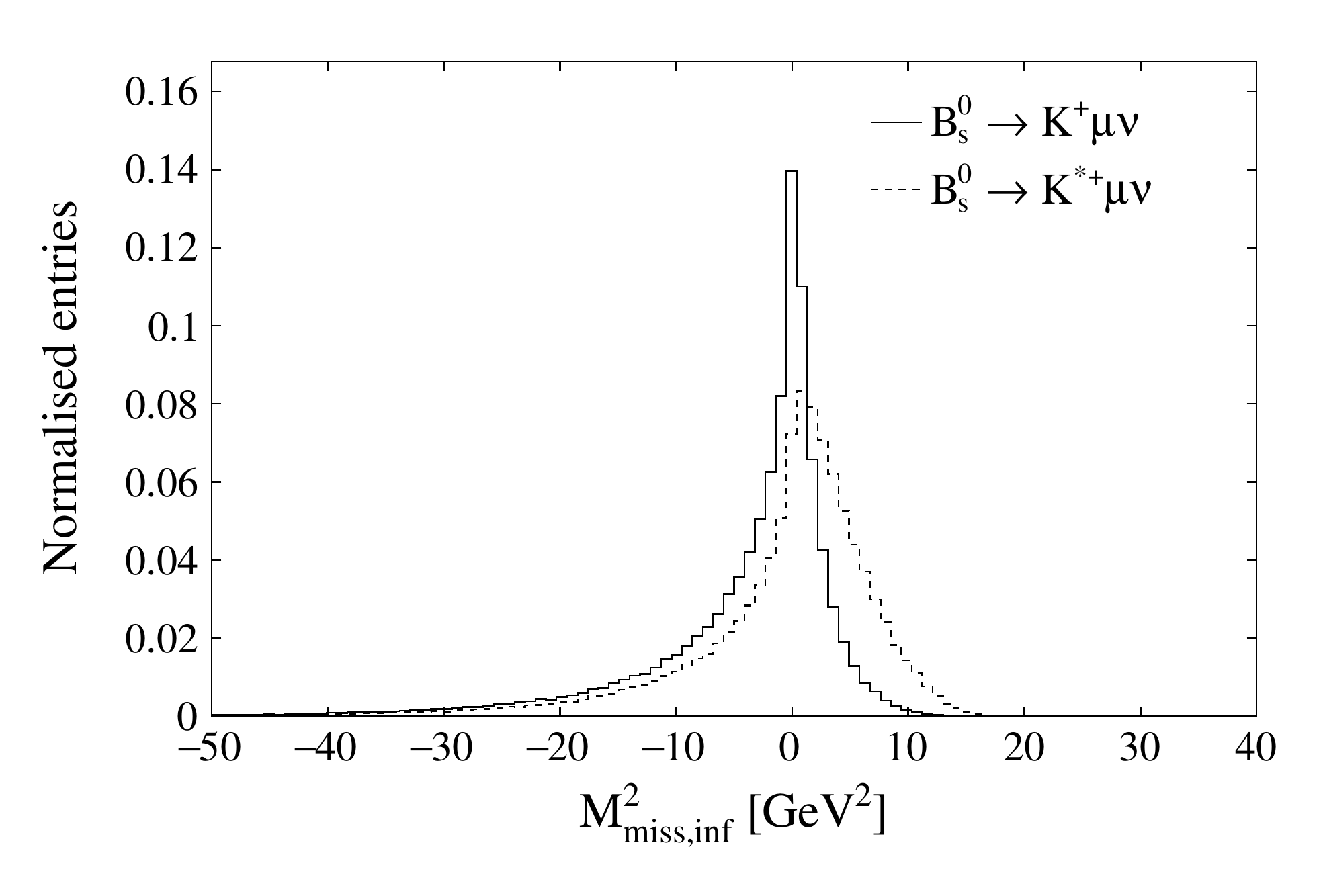}
\caption{\label{fig:1DSB}Comparison of various kinematic variables between simulated \Bs decays
to the $K\mu\nu$ and 
$K^{*}\mu\nu$ final states:
(top left) visible mass,
(top right) corrected mass,
(lower left) $b$-hadron mass using the 2-variable regression,
(lower right) squared missing mass using the 2-variable regression.}
\end{figure}

\begin{figure}[tb]\centering
\includegraphics[width=\widefigwidth\linewidth]{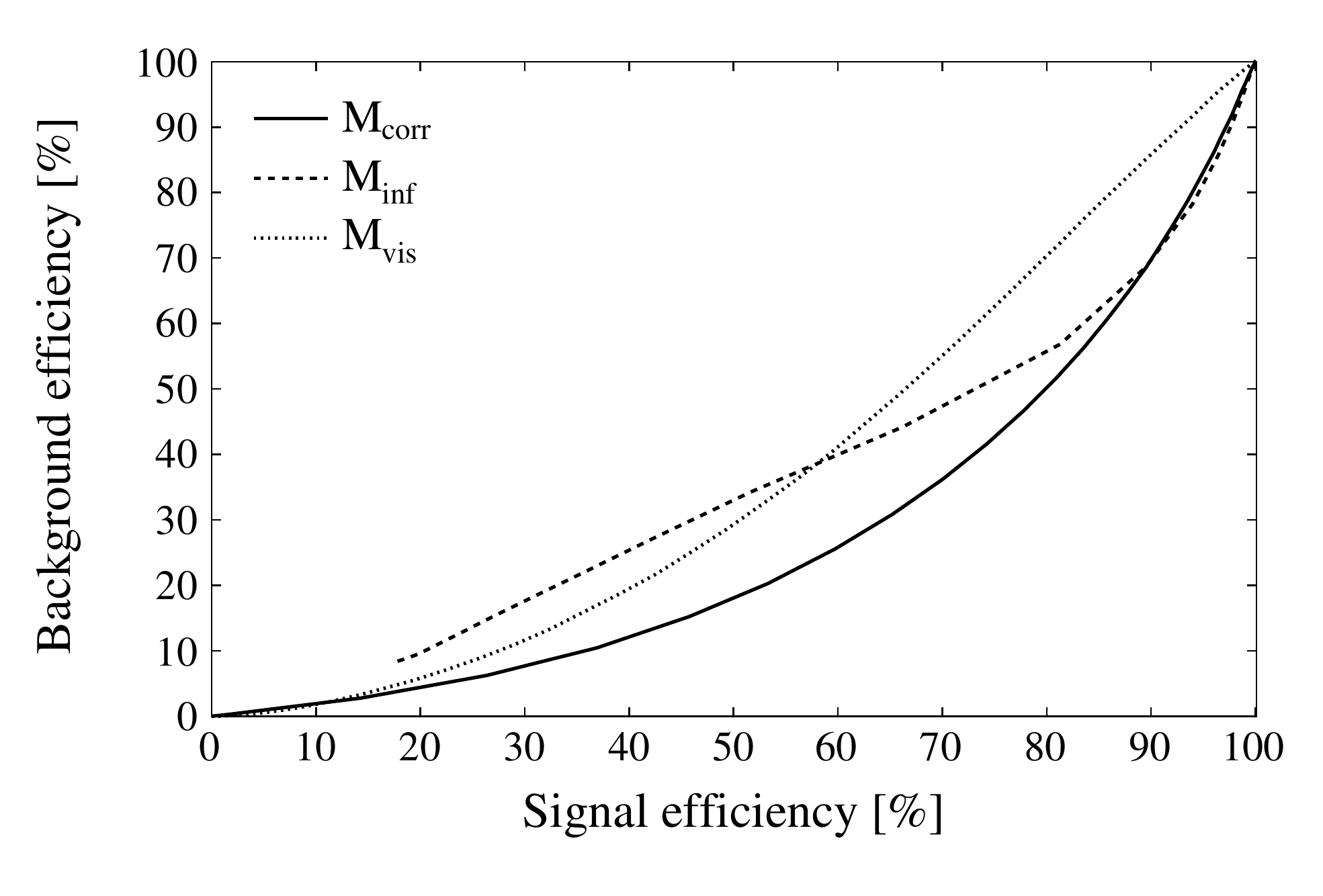}
\caption{\label{fig:ROC}The background (\BsToKstMuNu) efficiency versus the signal (\BsToKMuNu) efficiency for three different variables. }
\end{figure}

In this section we consider the use of \Pinf to define an optimal variable
for discriminating between decays with differing quantities of missing mass.
We take as an example the separation of 
\BsToKMuNu from \BsToKstMuNu, with $K^{^-} \to K^-\pi^0$, where the $\pi^0$
isn't reconstructed.
The corrected mass variable is defined with respect to the flight vector as~\cite{Abe:1997sb},
\begin{align}
\MCorr = \sqrt{\left(\MVis\right)^2 + \left(\PTMiss\right)^2 }+ \PTMiss.
\end{align}
It is heavily used in the LHCb trigger~\cite{LHCb-DP-2012-004} to inclusively select $b$-hadron decays,
and was the main discriminating variable that was used to extract the yield of 
\LbToPMuNu decays in the LHCb analysis of that mode~\cite{Aaij:2015bfa}.
The upper row of Fig.~\ref{fig:1DSB} shows the \MCorr and \MVis distributions
for the two \Bs decay modes.
The lower row shows two new variables that can be computed with the help of the regression based $b$-hadron momentum estimate.
In the first case it is assumed that the missing system has zero mass, which permits a computation of the parent $b$-hadron mass, denoted $M_{\rm inf}$. The distribution of this variable is shown in Fig.~\ref{fig:1DSB} (lower left).
Alternatively, the mass of the decaying $b$-hadron can be assumed, and a squared missing mass estimate 
can be made. The distribution of this variable, denoted $M^2_{\rm miss,inf}$,
is shown in Fig~\ref{fig:1DSB} (lower right).
The two new variables provide clear discrimination
but their performance should be compared to the established \MCorr variable.
Fig.~\ref{fig:ROC} shows the efficiency of \BsToKstMuNu versus the 
efficiency \BsToKMuNu for a range of cuts on \MCorr, \MVis and
$M_{\rm inf}$\footnote{Since the $b$-hadron mass and missing mass squared variables are essentially different transformations
of the same information, we only consider the $b$-hadron mass.}.
Interestingly, $M_{\rm inf}$ performs slightly better than $M_{\rm corr}$ in the region of signal efficiencies
in excess of 90\%.
However, \MCorr performs better in all other regions.

\subsection{Application to the study of b meson mixing and production}
\label{subsec:mixing}

Despite the obvious drawback of unreconstructible particles, 
semileptonic $b$-hadron decays have several notable advantages
for certain production and mixing studies.
Some of these benefit from the large signal yields in Cabibbo favoured
semileptonic decays and/or from the fact that these decays
are dominated by a single tree-level amplitude which 
limits direct \CP violation to a negligible level.
A recent review of \CP violation studies with \Bs mesons is provided by Ref.~\cite{Artuso:2015swg}.

\begin{figure}[tb]\centering
\includegraphics[width=\figwidth\linewidth]{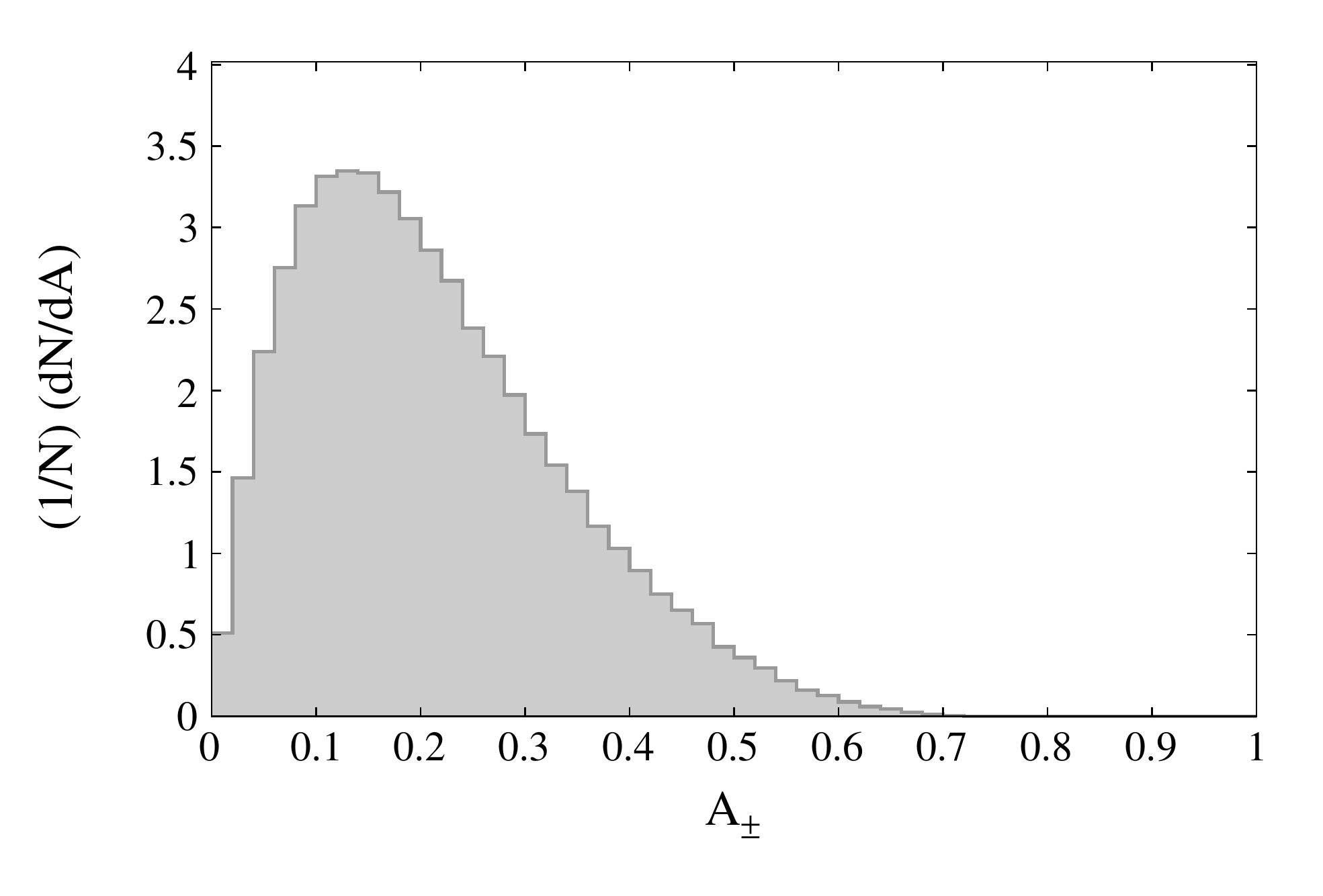}
\includegraphics[width=\figwidth\linewidth]{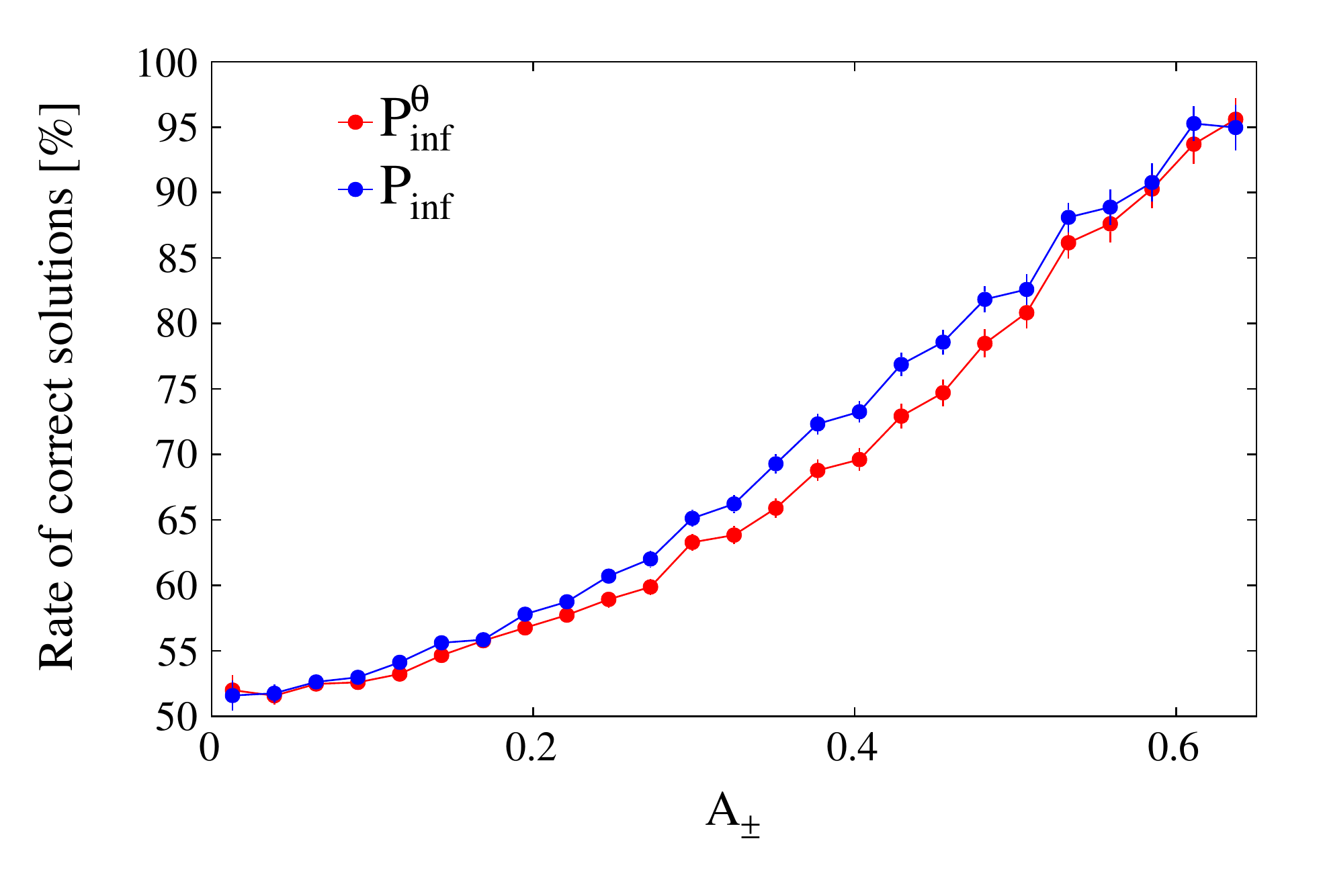}
\caption{\label{fig:DiffP2}Left: the \Apm distribution in simulated \BsToDsMuNu decays.
Right: the rate of correct solutions as a function of \Apm.}
\end{figure}

In Sect.~\ref{sec:Decays} it is noted that the correct solution rate
increases with $q^2$ in \BsToKMuNu decays because the two solutions become more distinct.
For the purpose of studying mixing and production it may be 
possible to build upon and exploit this feature.
We define the following asymmetry between the two solutions,
\begin{equation}
\label{eq:A}
\Apm = \frac{P_{+} - P_{-}}{P_{+}+P_{-}}.
\end{equation}
In Fig.~\ref{fig:DiffP2} the distribution of this variable 
in simulated \BsToDsMuNu decays is shown on the left, while 
on the right it can clearly be seen how the rate of correct solutions
increases as a function of \Apm.

In order to demonstrate the potential of our method we consider
the example of making a differential measurement
of the \BsBsb production asymmetry using \BsToDsMuNu decays.
This is highly challenging since 
it requires that the very fast 
\BsBsb oscillations can be resolved.
Semileptonic decays were recently used to make the single most 
precise measurement to date of the \BdBdb mixing frequency~\cite{Aaij:2016fdk}, but \BsBsb oscillations are
far more susceptible to the effects of missing momentum on the decay time resolution.
Nevertheless the first observation of \BsBsb oscillations by the CDF experiment
was based on a combination of hadronic and semileptonic \Bs decays~\cite{Abulencia:2006ze},
and LHCb made an observation of this phenomenon using semileptonic \Bs decays alone~\cite{Aaij:2013gja}.
In these analyses, the effect of the mis-measured momentum was corrected for
on a statistical basis using a Monte-Carlo correction that scales the $b$-hadron
momentum by a factor that depends on its visible mass.
%Hadronic \Bs decays are better suited to the measurement of the \BsBsb
%mixing frequency but the semileptonic mode \BsToDsMuNu is essential in the study of \CP violation.
The asymmetry between the number of \BsToDsMuNu and \BsbToDsMuNu decays, as a function of decay time,
can be written,
\begin{equation}
\label{Eq:asl}
\frac{\Gamma[D_s^-\mu^+,t] - \Gamma[D_s^+\mu^-,t]}{\Gamma[D_s^-\mu^+,t] + \Gamma[D_s^+\mu^-,t]} = \frac{a_{\rm sl}^s}{2} - \left[\frac{a_{\rm sl}^s + 2A_P}{2}\right] \left[\frac{\cos(\Delta M_s t)}{\cosh(\Delta \Gamma_s t /2)}\right],
\end{equation}
where $\Delta\Gamma_s$ is the decay width difference between the eigenstates of the \BsBsb system, $a_{\rm sl}^s$ is a \CP violating asymmetry and $A_P$ is the asymmetry in the rate of \Bs versus $\bar{B}_s$ production.
LHCb has measured~\cite{Aaij:2013gja,LHCb-PAPER-2014-053} $a_{\rm sl}^s$ using a decay time integrated version of Eq.~\ref{Eq:asl}, in which the second term can be shown to vanish~\cite{LHCb-PAPER-2014-053}.
It was recently suggested to study the decay time dependence in order to simultaneously
measure $a_{\rm sl}^s$ and $A_P$~\cite{Fleischer:2016dqd}\footnote{The authors also consider the application to a measurement of the \CP asymmetry in $D_s^{\pm}$ meson decays.}.
The production asymmetry is of interest in its own right
and a differential study as a function of rapidity ($y$) would be 
desirable~\cite{Norrbin:2000zc,Norrbin:2000jy,Norrbin:1999by},
 but subject to the resolution in the energy and momentum of the \Bs
which enter the definition of $y$.
It should also be noted that it would be experimentally challenging to 
control the feed-down from \BsToDstMuNu  decays
for which Eq.~\ref{eq:PpPm} is spoilt by the continuous spectrum of missing mass.
A statistical separation of these two components should be possible,
for example using the corrected mass variable,
though a fully realistic analysis is beyond the scope of the present study.

\begin{figure}[tb]\centering
\includegraphics[width=\figwidth\linewidth]{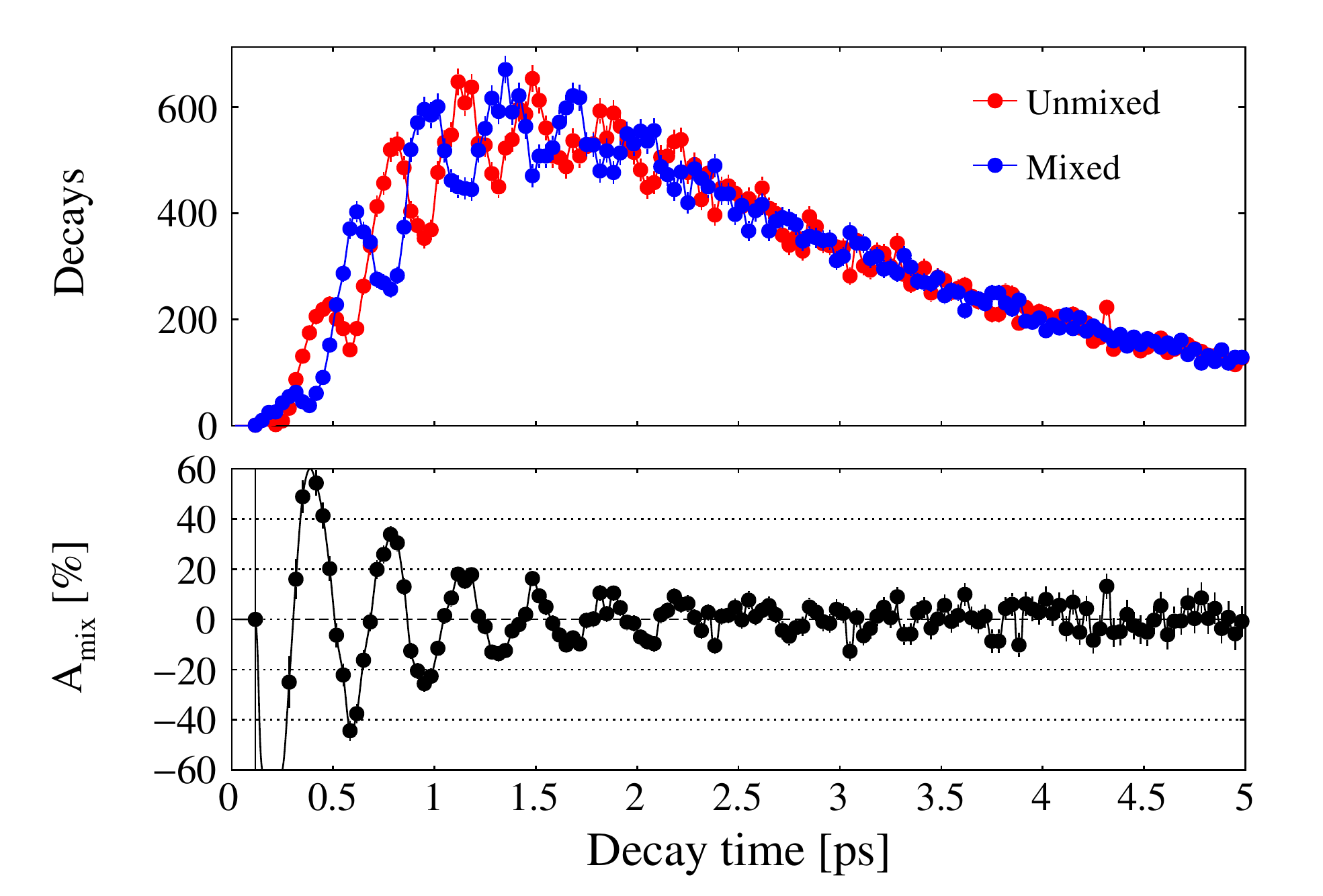}
\includegraphics[width=\figwidth\linewidth]{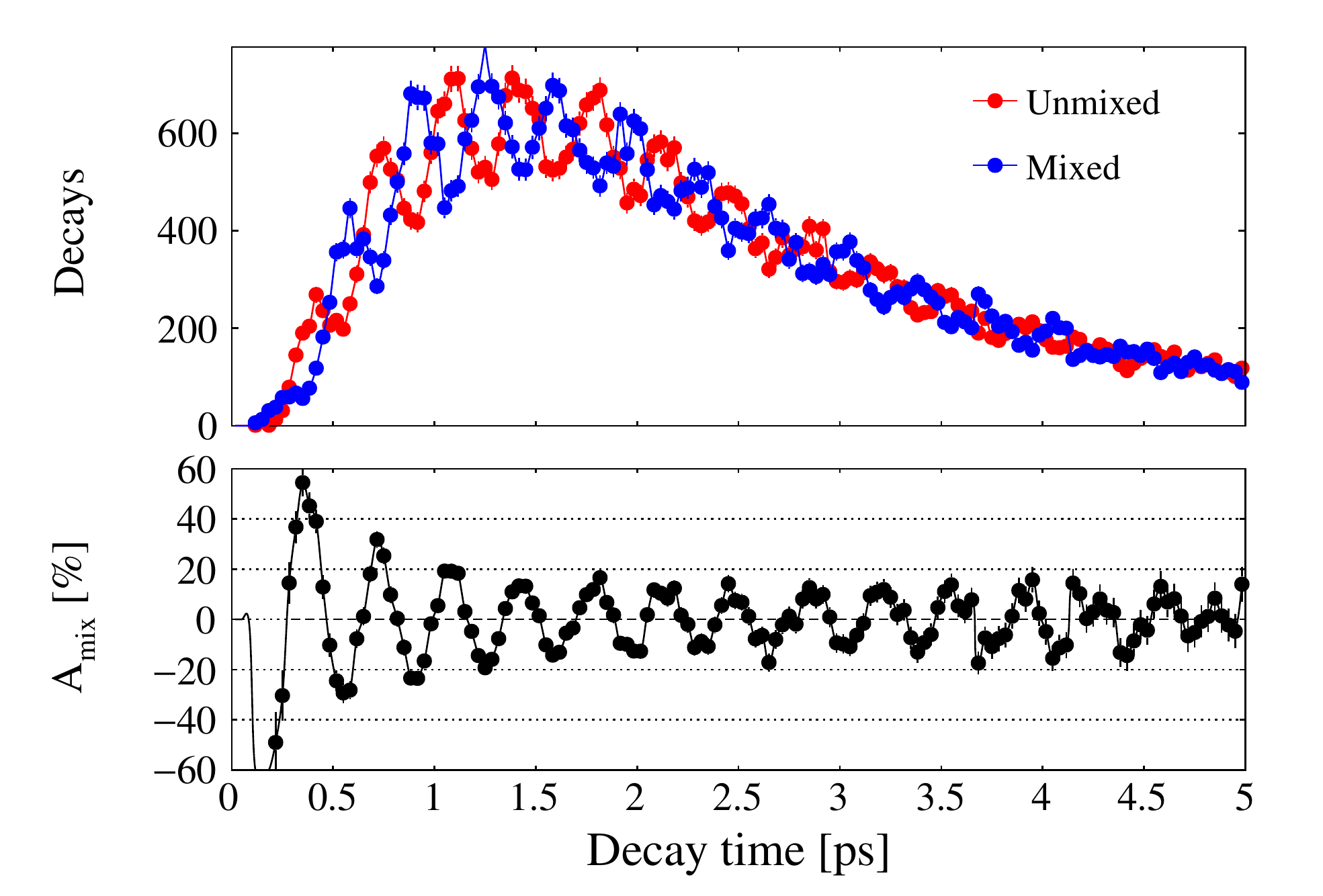}
\includegraphics[width=\figwidth\linewidth]{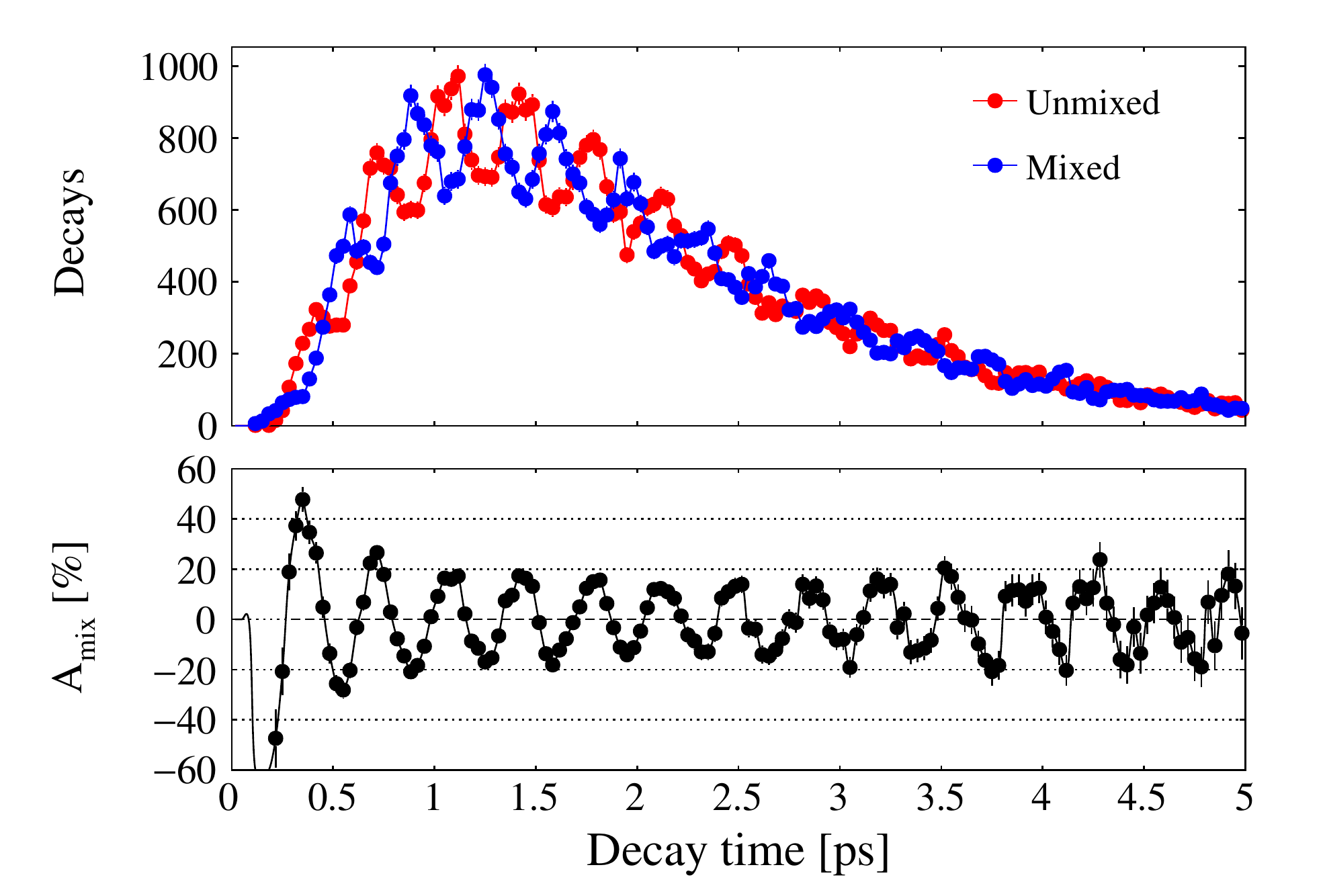}
\includegraphics[width=\figwidth\linewidth]{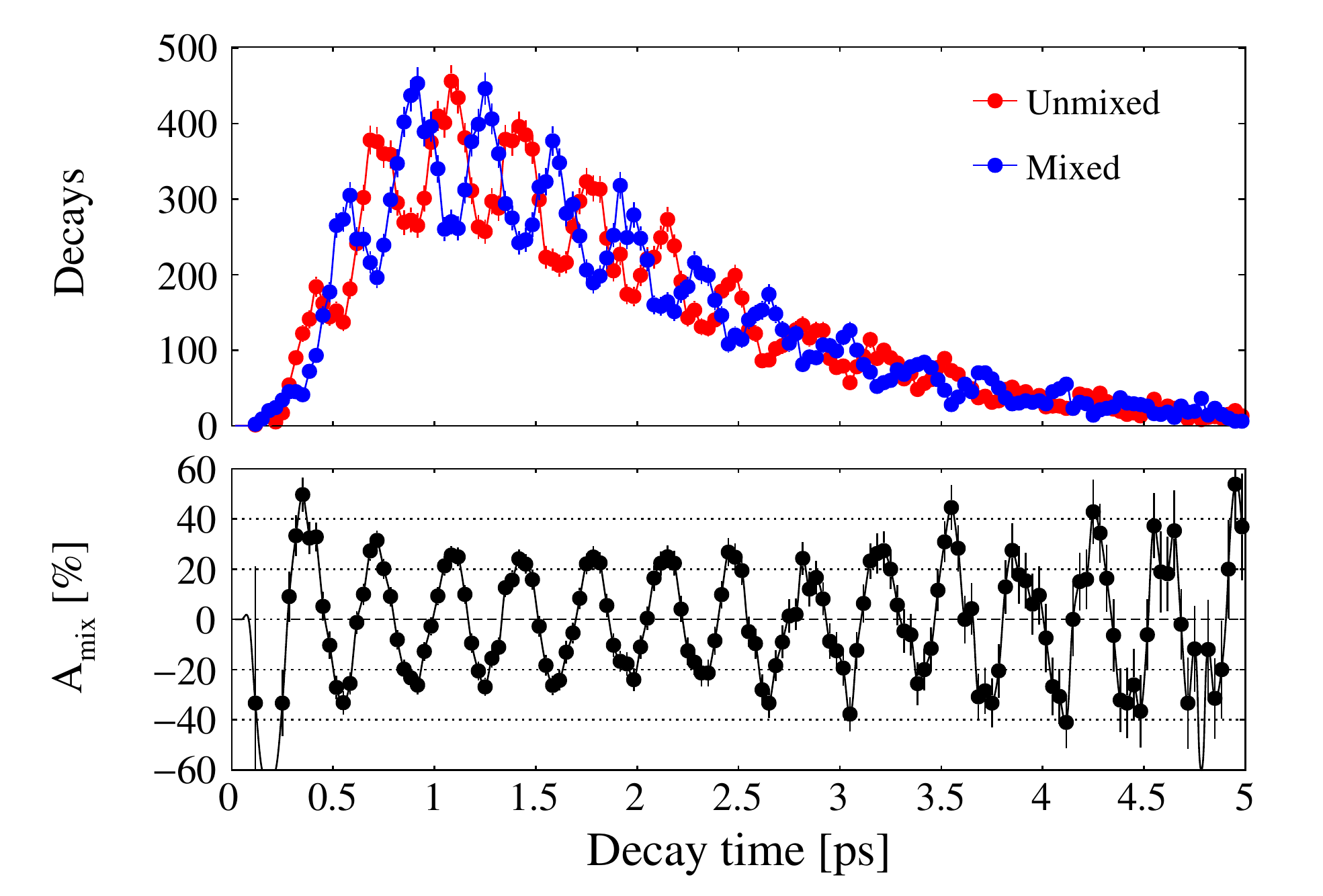}
\caption{\label{fig:Mixing}In each figure, the upper panel shows the decay time
distribution, separately for decays that are tagged as mixed and unmixed.
The lower panel shows the asymmetry between the rates of mixed and unmixed decays.
The first three figures correspond to different knowledge of the $b$-hadron momentum in the 
decay time computation: (upper left) raw visible momentum, (upper right)
random choice of quadratic solution, (lower left) regression based best quadratic solution.
On the lower right, the best quadratic solution  is used, 
and $\Apm > 0.23$ is required.}
\end{figure}

In order to study our ability to resolve the fast \Bs oscillations, we consider 
the decay time dependent mixing asymmetry defined as,
\begin{equation}
\label{Eq:mix}
A_{\rm mix}(t) \equiv \frac{N_{\rm mixed}(t) - N_{\rm unmixed}(t)}{N_{\rm mixed}(t) + N_{\rm unmixed}(t)},
\end{equation}
where $N_{\rm mixed}$ and $N_{\rm unmixed}$ are the numbers of reconstructed decays
that are tagged as mixed and unmixed, respectively.
LHCb currently achieves an effective tagging power of a few percent~\cite{Aaij:2016psi}
but this hypothetical study assumes perfect tagging.
Fig.~\ref{fig:Mixing} shows the mixed and unmixed decay time distributions and $A_{\rm mix}$
for various different measures of the decay time.
Using the uncorrected visible momentum, the first oscillation is well resolved,
but subsequent oscillations are rapidly smeared out.
This behaviour is not improved when scaling the momentum by a visible mass dependent factor.
If a random solution to the quadratic equation is used, then the damping of the oscillations
is slower.
Applying our method to select the best solution the oscillations are resolved
with an amplitude of around 20\% with minimal degradation at higher decay times.
Fig.~\ref{fig:Mixing} (lower right) shows how the sensitivity to the oscillations can be enhanced
by restricting to a region of larger \Apm ($>$ 0.23).\footnote{It 
should be noted however that in the high asymmetry region 
it may be more challenging to disentangle the \BsToDsMuNu and \BsToDstMuNu components.}

%by splitting into higher and lower regions of \Apm as defined in Eq.~\ref{eq:A}.

\begin{figure}[tb]\centering
\includegraphics[width=\figwidth\linewidth]{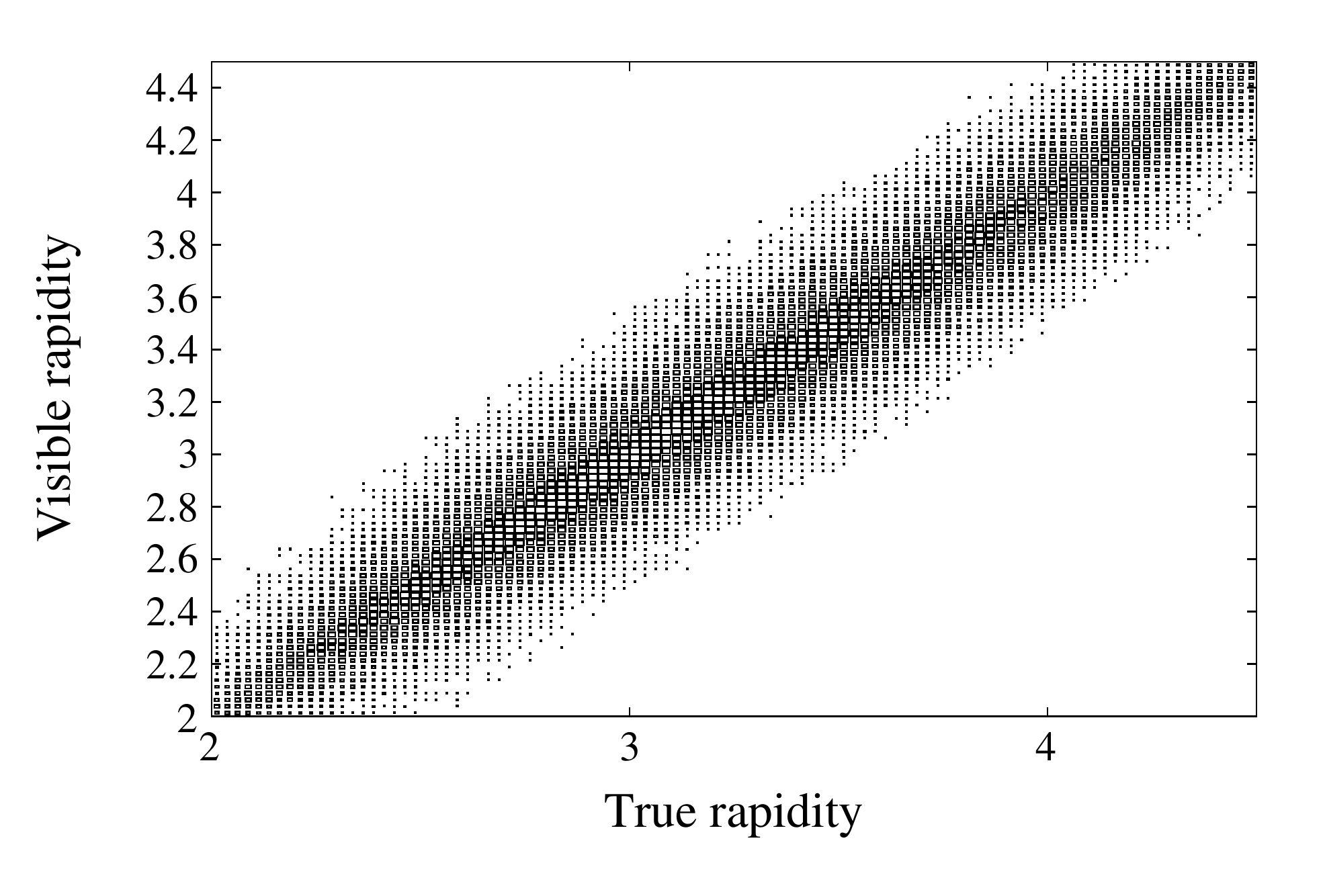}
\includegraphics[width=\figwidth\linewidth]{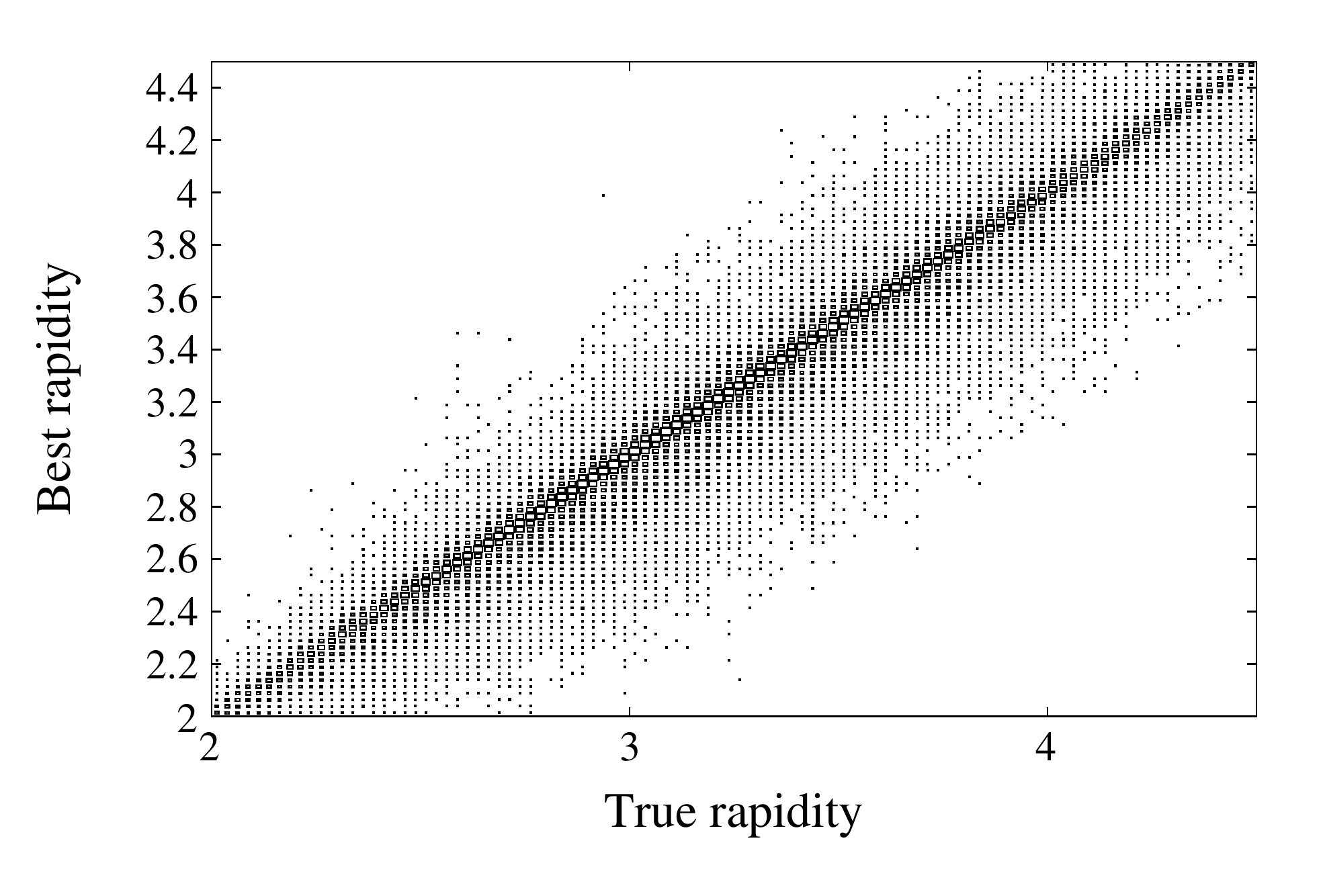}
\caption{\label{fig:Yres2D}The distributions of (left) the visible rapidity and (right) the 
best quadratic rapidity, versus the true rapidity.}
\end{figure}

\begin{figure}[tb]\centering
\includegraphics[width=\figwidth\linewidth]{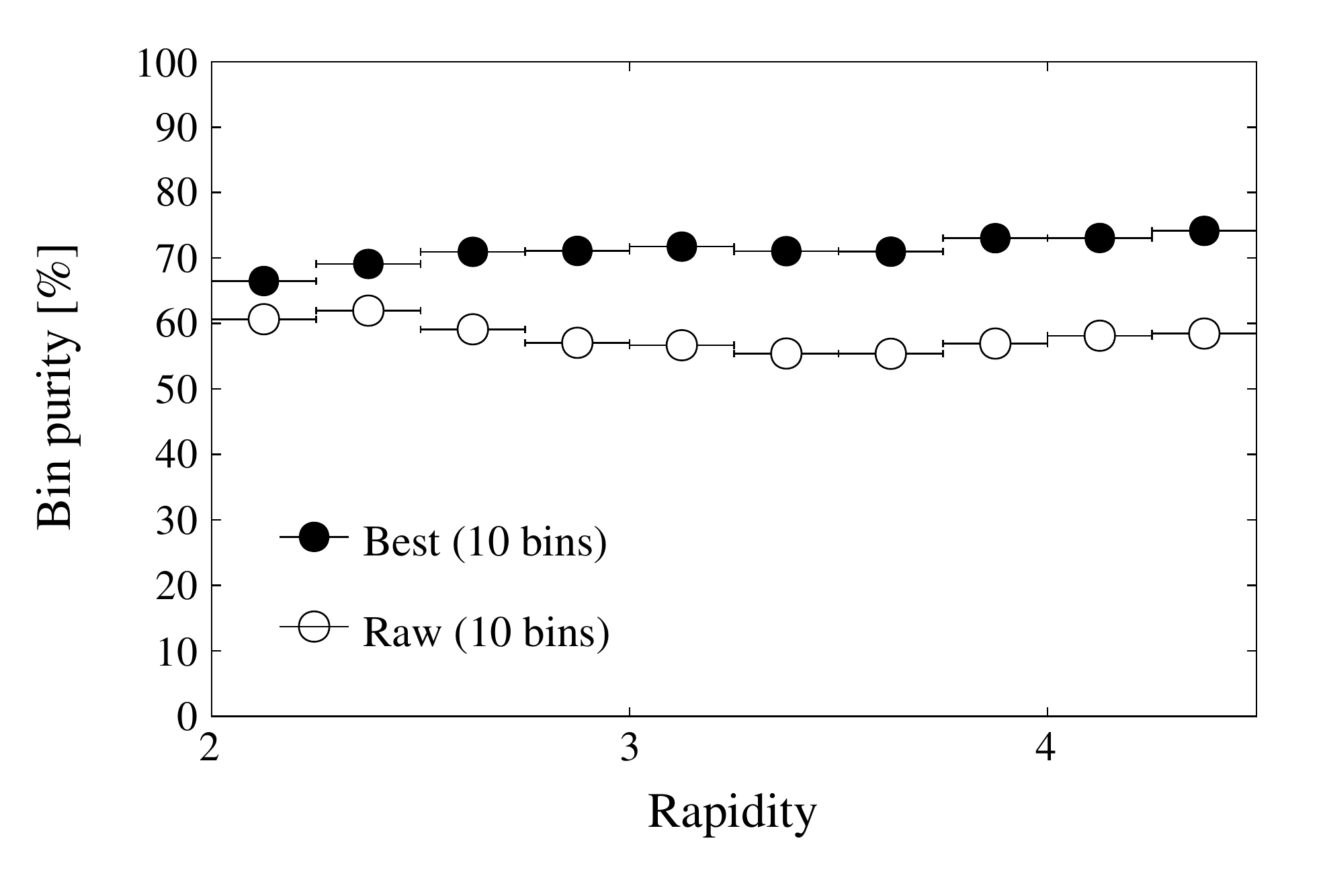}
\includegraphics[width=\figwidth\linewidth]{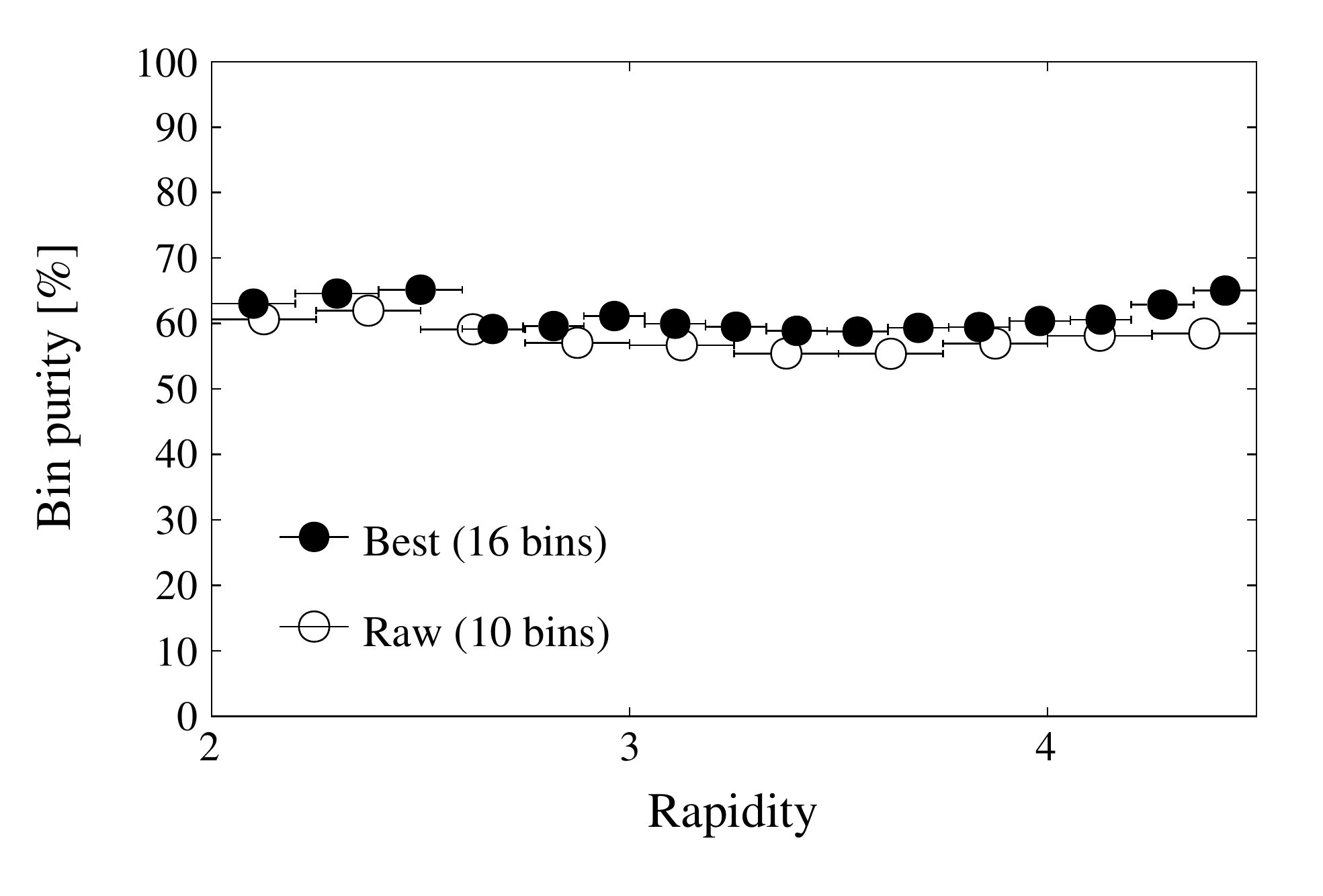}
\caption{\label{fig:Ypurity}The bin purity in rapidity for two different binning schemes.
The open points correspond to the visible rapidity.
The closed points correspond to the best quadratic rapidity.}
\end{figure}

The next challenge is to perform this analysis in 
bins of rapidity while unfolding for resolution.
Fig.~\ref{fig:Yres2D} (left) shows the distribution of the visible rapidity
versus the true rapidity. 
It can be seen in Fig.~\ref{fig:Yres2D} (right) that the corresponding
distribution with the best quadratic rapidity shows a better resolution
in this variable.
Fig.~\ref{fig:Ypurity} shows the bin purities in $y$
for different choices of bin boundaries.
A comparison is made for (open markers) $y$ defined using the
visible energy and momentum and (closed markers) $y$ defined
using the best quadratic solution with our method.
It is clear that the later approach will permit a finer binning.
While the mixing element of the example analysis with \Bs mesons may prove to be very challenging, 
the improved rapidity definition is applicable to production studies
with other $b$-hadron species for which a more straightforward decay time
integrated study is sufficient.

%\begin{figure}[tb]\centering
%\includegraphics[width=\figwidth\linewidth]{Fig16a.pdf}
%\includegraphics[width=\figwidth\linewidth]{Fig16b.pdf}
%\caption{\label{fig:MixingDiffA}The mixing asymmetry in the simulated \BsToDsMuNu decays
%roughly split into equal populations with (left) $A < 0.16$ and (right) $A> 0.16$.}
%\end{figure}

\section{Conclusions}

Hadron collider experiments offer a unique opportunity to study
semileptonic decays of $b$-hadron species that cannot be produced at $e^+e^-$
colliders operating at the $\Upsilon(4S)$ resonance.
The presence of unreconstructible final state particles poses a challenge
at hadron colliders since the momenta of the $b$-hadrons are poorly defined.
It is well known that in decays with a single missing particle, the decay 
kinematics can be reconstructed up to a quadratic ambiguity, by assuming the 
mass of the decaying $b$-hadron and imposing 
momentum conservation with respect to the reconstructed line of flight between
the production and decay vertices.
An independent estimate of the $b$-hadron momentum would be valuable since it 
could resolve the quadratic ambiguity.
We propose a method with which to estimate the momenta of the $b$-hadrons using
information that is totally independent of the $b$-hadron decays.
This exploits the fact that the flight direction and decay length of a $b$-hadron
are well measured and are both correlated to the $b$-hadron momentum.
A simple regression analysis yields a resolution of around 60\% in the $b$-hadron momentum.
This seemingly modest momentum estimate is sufficient to 
select the correct solution to the quadratic equation with an average 
rate of around 70\%.
This means that differential measurements of decay rates as a function of 
the invariant mass squared of the lepton-neutrino system in semileptonic decays
can be made with finer binning.
The method can also be applied in studies of $b$-hadron production.
As a highly challenging test case we consider the example of measuring 
the \BsBsb production asymmetry in bins of rapidity, which requires
the resolution of the fast \BsBsb oscillations.
Our method permits a finer binning in rapidity and improves the resolution on the \BsBsb oscillations.
The general method can be perfectly validated using control samples of fully reconstructed $b$-hadron decays,
and it should be easy to apply to analyses since it only requires knowledge of the position of
the primary $pp$ interaction vertex and the $b$-hadron decay vertex.

\acknowledgments

We thank Vladimir Gligorov, Suzanne Klaver, Patrick Owen, Konstantinos Petridis, Nicola Serra and Vincenzo Vagnoni for interesting discussions
and suggestions during the preparation of this manuscript.
M.~V.~acknowledges support from the Alexander von Humboldt foundation.

%\paragraph{Note added.} This is also a good position for notes added
%after the paper has been written.

% The bibliography will probably be heavily edited during typesetting.
% We'll parse it and, using the arxiv number or the journal data, will
% query inspire, trying to verify the data (this will probalby spot
% eventual typos) and retrive the document DOI and eventual errata.
% We however suggest to always provide author, title and journal data:
% in short all the informations that clearly identify a document.

\addcontentsline{toc}{section}{References}
\setboolean{inbibliography}{true}
\bibliographystyle{LHCb}
\bibliography{main}

\end{document}